\documentclass[10pt,journal,twocolumn]{IEEEtran}

\usepackage[utf8]{inputenc}
\usepackage{graphicx}
\usepackage{stfloats}
\usepackage{amssymb}
\usepackage[cmex10]{amsmath}
\usepackage{color}
\usepackage{pifont}
\usepackage{hyperref}
\usepackage{cite}
\usepackage{url}
\usepackage{breakurl}
\usepackage{bbm}
\usepackage[ruled,vlined,titlenumbered]{algorithm2e}
\usepackage{mathdots}
\usepackage{capt-of}
\usepackage{subfigure}
\usepackage{ifthen}

\newcommand{\F}{\ensuremath{\mathbb{F}}}
\newcommand{\Fx}{\ensuremath{\mathbb{F}[x]}}
\newcommand{\GF}[1]{\text{GF}($#1$)}
\newcommand{\Fxsub}[1]{\ensuremath{\mathbb{F}_{#1}[x]}}
\newcommand{\Fxy}{\ensuremath{\mathbb{F}[x,y]}}
\newcommand{\N}{\ensuremath{\mathbb{N}}}

\newcommand{\EKE}[1]{\ensuremath{\operatorname{EKE}(#1)}}
\newcommand{\EKEzero}[1]{\ensuremath{\operatorname{EKE}_0(#1)}}
\newcommand{\refeq}[1]{(\ref{#1})}
\newcommand{\wdeg}[2]{\ensuremath{\operatorname{wdeg_{#1,#2}}}}
\newcommand{\ev}{\ensuremath{\operatorname{ev}}}

\newtheorem{definition}{Definition}
\newtheorem{theorem}{Theorem}
\newtheorem{lemma}{Lemma}
\newtheorem{proposition}{Proposition}
\newtheorem{problem}{Problem}

\newtheorem{note}{Note}

\newcommand{\UHASSE}[1]{\ensuremath{Q^{[#1]}}}
\newcommand{\SYNDPOL}[1]{\ensuremath{S^{(#1)}(x)}} 
\newcommand{\SYNDPOLb}[1]{\ensuremath{{T}^{(#1)}(x)}}
\newcommand{\SYNDCOEFF}[2]{\ensuremath{S^{(#1)}_{#2}}}
\newcommand{\mult}[2]{\ensuremath{\operatorname{mult}\big({Q^{#1}(x,y),(#2)}\big)}}
\newcommand{\multgen}[2]{\ensuremath{\operatorname{mult}\big(#1,(#2)}\big)}

\newcommand{\LCOEFF}[2]{\ensuremath{\Lambda^{(#1)}_{#2}}}
\newcommand{\LPOL}[1]{\ensuremath{\Lambda^{(#1)}(x)}}
\newcommand{\QPOL}[1]{\ensuremath{Q^{(#1)}(x)}}

\newcommand{\listl}{\ensuremath{\ell}}

\newcommand{\GRS}[2]{\ensuremath{\mathcal{GRS}(#1,#2)}}
\newcommand{\ON}[1]{\ensuremath{\mathcal{O}\left({#1}\right)}}
\newcommand{\printalgo}[1]
{
\scalebox{0.9}{
\centering
\begin{algorithm}[H]
 #1
\end{algorithm}
}
}

\newcommand{\IP}[2]{\ensuremath{\langle #1,#2 \rangle}}
\newcommand{\SET}[1]{\ensuremath{[#1]}}
\newcommand{\SETzero}[1]{\ensuremath{[#1]_0}}

\newcommand{\mat}[2][\empty]{
  \ifthenelse{\equal{#1}{\empty}}
    {\ensuremath{\mathbf{#2}}}
    {\ensuremath{#2_{#1}}}
}

\newcommand{\vect}[2][\empty]{
  \ifthenelse{\equal{#1}{\empty}}
    {\ensuremath{\mathbf{#2}}}
    {\ensuremath{#2_{#1}}}
}

\newcommand{\dhalf}{\ensuremath{\left\lfloor (n-k)/2\right\rfloor}}
\DeclareMathOperator{\defi}{def}
\newcommand{\defeq}{\overset{\defi}{=}}

\newcommand{\nexth}[2]{\mbox{\ensuremath{\operatorname{succ} ( \prec_H,(#1,#2)) }}}
\newcommand{\nextv}[2]{\mbox{\ensuremath{\operatorname{succ} ( \prec_V,(#1,#2)) }}}
\newcommand{\prevv}[2]{\mbox{\ensuremath{\operatorname{pred} ( \prec_V,(#1,#2)) }}}

\newcommand{\noshow}[1]{}

\begin{document}
\title{An Interpolation Procedure for\\ List Decoding Reed--Solomon Codes\\ Based on Generalized Key Equations}

\author{{Alexander Zeh, Christian Gentner and Daniel Augot
\thanks{Parts of this work were published in the proceedings of 2008 IEEE International Symposium on Information Theory (ISIT 2008), Toronto, Canada~\cite{AugotZeh_OnTheRREquations_2008} and  2009 IEEE Information Theory Workshop (ITW 2009), Taormina, Sicily, Italy~\cite{Zeh_EfficientListDecoding_2009}.
This work was supported by the German Research Council "Deutsche Forschungsgemeinschaft" (DFG) under Grant No. Bo867/22-1.
}}}

\maketitle

\begin{abstract}
The key step of syndrome-based decoding of Reed--Solomon codes up to
half the minimum distance is to solve the so-called Key Equation. 
List decoding algorithms, capable of decoding beyond half
the minimum distance, are based on interpolation and factorization of multivariate polynomials. 
This article provides a link between
syndrome-based decoding approaches based on Key Equations and the
interpolation-based list decoding algorithms of Guruswami and Sudan for Reed--Solomon codes.  
The original interpolation conditions of Guruswami and Sudan for Reed--Solomon codes are reformulated in terms
of a set of Key Equations. These equations provide a structured
homogeneous linear system of equations of Block-Hankel form, that can be solved
by an adaption of the Fundamental Iterative Algorithm. For
an $(n,k)$ Reed--Solomon code, a multiplicity $s$ and a list size $\listl$, our
algorithm has time complexity \ON{\listl s^4n^2}.
\end{abstract}

\begin{IEEEkeywords}
Guruswami--Sudan interpolation, Key Equation, list decoding, Fundamental Iterative Algorithm (FIA), Reed--Solomon codes,
Hankel matrix, Block-Hankel matrix.
\end{IEEEkeywords}

\newcounter{mytempeqncnt}
\newcounter{mytempeqncntstore}

\section{Introduction}

\IEEEPARstart{I}{n} 1999, Guruswami and
Sudan~\cite{GuruswamiSudan_ImproveddecodingofReed-Solomonandalgebraic-geometrycodes_1999,Guruswami_ListDecodingofError-CorrectingCodes_1999,Guruswami_ALGORITHMICRESULTSINLISTDECODING_2007}
extended Sudan's original approach~\cite{sudan97decoding} by introducing multiplicities
in the interpolation step of their polynomial-time list decoding
procedure for Reed--Solomon and Algebraic Geometric codes. This
modification permits decoding of $(n,k)$ Reed--Solomon
codes~\cite{ReedSolomon_PolynomialCodesOverCertainFiniteFields_1960} (and Algebraic Geometric codes) of arbitrary code-rate $R=k/n$
with increased decoding radius. Guruswami and Sudan were focused
on the existence of a polynomial-time algorithm. Kötter~\cite{Koetter:THS1996} and Roth--Ruckenstein~\cite{Roth_Ruckenstein_2000,
Ruckenstein_PHD2001} proposed quadratic time algorithms for the key steps of the Guruswami--Sudan
principle for Reed--Solomon codes, i.e., interpolation and factorization of bivariate polynomials. Various other
approaches for a low-complexity realization of Guruswami--Sudan
exist, e.g. the work of
Alekhnovich~\cite{Alekhnovich_LinearDiophantineEquation_2005},
where fast computer algebra techniques are used. Trifonov's~\cite{Trifonov_EfficientInterpolation_2010}
contributions rely on ideal theory and divide and conquer
methods. Sakata uses Gr\"obner--bases techniques~\cite{Sakata_FindingAMinimalSet_1991, Sakata_OnFastInterpolation_2001}.

In this paper, we reformulate the \textit{bivariate} interpolation step of Guruswami--Sudan for Reed--Solomon
codes in a set of \textit{univariate} Key Equations~\cite{AugotZeh_OnTheRREquations_2008}. This extends the
previous work of Roth and Ruckenstein~\cite{Roth_Ruckenstein_2000,Ruckenstein_PHD2001}, where the
reformulation was done for the special case of Sudan. Furthermore, we
present a modification of the so-called Fundamental Iterative
Algorithm (FIA), proposed by Feng and Tzeng in
1991~\cite{Feng_Tzeng1991}. Adjusted to the special case of
one Hankel matrix the FIA resembles the approach of Berlekamp and Massey~\cite{Berlekamp:AGT1968, Massey_Shift-registersynthesisandBCHdecoding_1969}.

Independently of our contribution, Beelen and H{\o}holdt reformulated
the Guruswami--Sudan constraints for Algebraic Geometric codes~\cite{Beelen_ASyndromeFormulation_2008, Beelen_KeyEquations_2010}. 
It is not clear, if the system they obtain is highly structured.

This contribution is structured as follows. The next section contains
basic definitions for Reed--Solomon codes and bivariate polynomials.
In Section~\ref{sec_WBandFIA}, we derive the Key Equation for
conventional decoding of Reed--Solomon codes from the Welch--Berlekamp approach~\cite{BerlekampWelch_Patent} and we 
present the adjustment of the FIA for one Hankel matrix. A modified version of Sudan's reformulated interpolation
problem based on the work of Roth--Ruckenstein~\cite{Roth_Ruckenstein_2000} is derived and the
adjustment of the FIA for this case is illustrated in
Section~\ref{sec_horizontalline}. In
Section~\ref{sec_GSAndHankelMatrices}, the interpolation step of the Guruswami--Sudan
principle is reformulated. The obtained homogeneous set of linear equations has
Block-Hankel structure. We adjust the FIA for this Block-Hankel
structure, prove the correctness of the proposed algorithm and analyze its
complexity. We conclude this contribution in Section~\ref{sec_conclusion}.


\section{Definitions and Preliminaries} \label{sec_defs}

Throughout this paper, \SET{n} denotes the set of integers
$\{1,2,\dots,n\}$ and \SETzero{n} denotes the set of integers
$\{0,1,\dots,n-1\}$. An $m \times n$ matrix $\mat{A}=
\vert \vert \mat[i,j]{A} \vert \vert $ consists of the entries $\mat[i,j]{A}$, where
$i \in \SETzero{m}$ and $j \in \SETzero{n}$. A univariate polynomial $A(x)$ of degree less than $n$ is
denoted by $A(x) = \sum_{i=0}^{n-1} A_i x^i$. A vector of length $n$ is represented by
$\vect{r} = (\vect[1]{r},\vect[2]{r},\dots,\vect[n]{r})^T$.

Let $q$ be a power of a prime and let $\F=\GF{q}$ denote the finite
field of order $q$.  Let $\alpha_1, \alpha_2, \dots, \alpha_n $ denote
nonzero distinct elements (code--locators) of $\F$ and let 
$\upsilon_1^{\prime},\upsilon_2^{\prime},\dots, \upsilon_n^{\prime}$ denote nonzero elements (column--multipliers), the associated evaluation map
$\ev$ is
\begin{equation} \label{eq_evmap}
\ev:\begin{array}[t]{ccc}
  \F[x]& \rightarrow & \F^n\\
    f(x)& \mapsto & \left(\upsilon_1^{\prime}f(\alpha_1),\upsilon_2^{\prime}f(\alpha_2), \dots,\upsilon_n^{\prime}f(\alpha_n) \right).
\end{array}
\end{equation}
The associated Generalized Reed--Solomon code \GRS{n}{k} of length
$n$ and dimension $k$ is~\cite{MacWilliamsSloane_TheTheoryOfErrorCorrecting_1988}:
\begin{equation} \label{eq_defRS}
 \GRS{n}{k} = \{ \mathbf{c} = \ev(f(x)): f(x) \in \Fxsub{k} \},
\end{equation}
where $\Fxsub{k}$ denotes the set of all univariate polynomials with
degree less than $k$. Generalized Reed--Solomon codes are MDS codes
with minimum distance $d = n-k+1$.  The dual of a Generalized
Reed--Solomon is also a Generalized Reed--Solomon code with the same
code locators and column multipliers $\upsilon_1,\upsilon_2,\dots,\upsilon_n$, 
where $\sum_{i=1}^n \upsilon_i^{\prime}
\upsilon_i \alpha_i^j = 0,\; j \in
\SETzero{n-1}$. The explicit form of the column multipliers is~\cite{Roth_IntroductiontoCodingTheory_2006}:
\begin{equation} \label{eq_columnGRS}
\upsilon_i=\frac1{\upsilon^\prime_i}\cdot\frac1{\prod_{j\neq i}(\alpha_i-\alpha_j)}, \quad i\in\SET{n}.
\end{equation}

We will take advantage of structured matrices and therefore we recall
the definition of a Hankel matrix in the following.

\begin{definition}[Hankel Matrix] \label{def_Hankel}
 An $m \times n$ Hankel matrix $\mat{S} = \vert \vert \mat[i,j]{S} \vert \vert $ is a matrix, where $\mat[i,j]{S} = \mat[i-1,j+1]{S}$ for all $i
 \in \SET{m-1}$ and $j \in \SETzero{n-1} $ holds.
\end{definition}

Let us recall some properties of bivariate polynomials in $\Fxy$.

\begin{definition}[Weighted Degree] \label{def_wdeg}
Let the polynomial $A(x,y) = \sum_{i,j} A^{(j)}_i x^i y^j$ be in $\Fxy$.
Then, the $(w_x,w_y)$-weighted degree of $A(x,y)$, denoted by \wdeg{w_x}{w_y},
is the maximum over all $iw_x+jw_y$ such that $A^{(j)}_i \neq 0$.
\end{definition}

\begin{definition}[Multiplicity and Hasse Derivative~\cite{Hasse_1936}] \label{def_multiplicity}
Let $A(x,y) = \sum_{i,j} A^{(j)}_i x^i y^j$ be a polynomial in $\Fxy$.
Let $\overline{A}(x,y) = A(x+\alpha,y+\beta) = \sum_{i,j} \overline{A}^{(j)}_i x^i y^j$. 
A bivariate polynomial $A(x,y)$ has at least multiplicity $s$ in the point $(\alpha,\beta)$, denoted
by
\begin{equation} \label{eq_multnot}
\multgen{A(x,y)}{\alpha,\beta} \geq s,
\end{equation}
if the coefficients $\overline{A}^{(j)}_i$ are zero for all $i+j<s$. Furthermore, the $(a,b)$th Hasse derivative of the polynomial $A(x,y)$ in the 
point $(\alpha,\beta)$ is
\begin{equation} \label{eq_HasseDer}
\begin{split}
 A^{[a,b]}(\alpha,\beta) & = \sum_{i \geq a, j \geq b} \binom{i}{a} \binom{j}{b} A^{(j)}_i \alpha^{i-a} \beta^{j-b} 
\end{split}
\end{equation}
Let $A^{[b]}(x,y) = A^{[0,b]}(x,y)$ denote the $b$th Hasse derivative of $A(x,y)$ with respect to the
variable $y$.
\end{definition}

We will use the inner product for bivariate polynomials to describe our algorithms.

\begin{definition}[Inner Product] \label{def_InnerProduct}
Let two polynomials $A(x,y)=\sum_{i,j} A_i^{(j)} x^i y^j$ and $B(x,y)=\sum_{i,j} B_i^{(j)} x^i y^j$ in $\Fxy$ be given.
The inner product \IP{A(x,y)}{B(x,y)} of $A(x,y)$ and $B(x,y) $ 
is defined by $ \sum_{i,j} A_{i}^{(j)} B_{i}^{(j)}$.
\end{definition}


\section{Welch--Berlekamp as List-One Decoder and the Fundamental Iterative Algorithm} \label{sec_WBandFIA}
\subsection{Syndrome-Based Decoding of Reed--Solomon Codes} \label{sec_RSCodes}
Let $\mathbf{e} = (e_1,e_2,\dots,e_n)$ denote the error word and let $\mathcal{J}$ be the set of error locations (that is $e_j \neq 0  \Leftrightarrow j \in \mathcal{J}$).
Let $\tau = \dhalf$. It is well-known that a \GRS{n}{k} code can recover uniquely any error pattern if and only if  $\lvert \mathcal{J} \rvert \leq \tau $. 
The $n-k$ syndrome coefficients $S_0,S_1,\dots,S_{n-k-1}$ depend only on the error word $\mathbf{e}$ and
the associated syndrome polynomial $S(x)$ is defined by~\cite{Roth_IntroductiontoCodingTheory_2006}:
\begin{equation*} 
 S(x) =\sum_{i=0}^{n-k-1}S_i x^i\equiv \sum_{j=1}^n \frac{e_j\upsilon_j}{1-\alpha_j x} \mod x^{n-k}.
\end{equation*}
The error-locator polynomial is $\Lambda(x)=\prod_{j\in \mathcal{J}}(1-\alpha_j x)$ and 
the error-evaluator polynomial $\Omega(x)$ is $\sum_{j\in\mathcal{J}}e_j\upsilon_j\prod_{i \in \mathcal{J}\setminus\{j\}}(1-\alpha_i x)$.
They are related by the Key Equation:
\begin{equation} \label{eq_KeyEquation_frac}
\frac{\Omega(x)}{\Lambda(x)}\equiv S(x)\bmod x^{n-k}.
\end{equation}
The main steps for conventional decoding up to half the minimum distance are:
\begin{enumerate}
 \item Calculate the syndrome polynomial $S(x)$ from the received word $\mathbf{r} = \mathbf{c}+\mathbf{e}$.
 \item Solve~\refeq{eq_KeyEquation_frac} for the error-locator polynomial $\Lambda(x)$ and determine its roots.
 \item Compute $\Omega(x)$ and then determine the error values.
\end{enumerate}

\subsection{Derivation of the Key Equation from Welch--Berlekamp}
We derive the classical Key Equation~\refeq{eq_KeyEquation_frac} from the simplest interpolation based decoding algorithm, reported as the
``Welch--Berlekamp'' decoding algorithm in~\cite{Gemmell-Sudan_IFL1992,YaghoobianBlake_TwoNewDecodingAlgorithmsForRSCodes_1994,DabiriBlake_FastParallelAlgorithmsRemainder_1995}. 
We provide a simpler representation than in~\cite{BerlekampWelch_Patent} and give a polynomial derivation of the Key Equation.

Consider a \GRS{n}{k} code with support set
$\alpha_1,\alpha_2,\dots,\alpha_n$, multipliers $\upsilon^{\prime}_1,\upsilon^{\prime}_2,\dots,\upsilon^{\prime}_n$ and
dimension $k$. The Welch--Berlekamp approach is based on the following
lemma~\cite[Ch. 5.2]{JustesenHoholdt_ACourseinError-CorrectingCodes_2004}:
\begin{lemma}[List-One Decoder] \label{lem_wbuni}
Let $\mathbf{c}= \ev(f(x))$ be a codeword of a \GRS{n}{k} code and let $\mathbf{r}=\mathbf{c}+\mathbf{e} = (r_1,r_2,\dots,r_n)$ be the received word.
We search for a polynomial $Q(x,y) = \QPOL{0} + \QPOL{1}y $  in $\Fxy$ such that:
\begin{enumerate}
\item $Q(x,y)\neq 0$,
\item $Q(\alpha_i,r_i/\upsilon^{\prime}_i)=0 \quad  \forall i \in \SET{n}$,
\item $\wdeg{1}{k-1}Q(x,y) < n-\tau $.
\end{enumerate}
If $\mathbf{c}$ has distance less than or equal to $\dhalf$ from the received word $\mathbf{r}$, then $f(x)= -\QPOL{0}/\QPOL{1}$.
\end{lemma}

Let us connect Lemma~\ref{lem_wbuni} to~\refeq{eq_KeyEquation_frac}.

\begin{proposition}[Univariate Reformulation] \label{prop_WelchBerlekamp}
Let $R(x)$ be the Lagrange interpolation polynomial, such that
$R(\alpha_i) = r_i /\upsilon^{\prime}_i, \;  i \in \SET{n}$ holds. Let
$G(x)=\prod_{i=1}^n(x-\alpha_i)$. Then $Q(x,y)=\QPOL{0} + \QPOL{1}y $ satisfies
Conditions \emph{2)} and \emph{3)} of Lemma~\ref{lem_wbuni} if and only if
there exists a polynomial $B(x) \, \in \, \Fx$ such that
\begin{equation} \label{eq_list1_start}
 Q(x,R(x)) = B(x) \cdot G(x),
\end{equation}
and $\deg B(x) < n-k-\tau$.
\end{proposition}
Let $N_t=n-\tau-t(k-1), \, t=0,1$. Define the following reciprocal polynomials:
\begin{equation} \label{eq_list1_reverse_1}
\begin{split}
 \overline R(x) & = x^{n-1} \cdot R(x^{-1}) ,  \\ 
 \overline G(x) & = x^n \cdot G(x^{-1}) = \prod_{i=1}^n(1-\alpha_ix), \\
 \Omega(x) & = x^{\deg B(x)} \cdot B(x^{-1}), \\
 \LPOL{t} & = x^{N_t-1} \cdot Q^{(t)}(x^{-1}).
\end{split}
\end{equation}
Inverting the order of the coefficients of~\refeq{eq_list1_start} leads to:
\begin{equation*}
 \begin{split}
  x^{n-\tau+n-k-1}  \big( Q^{(0)}(x^{-1})   + Q^{(1)}(x^{-1}) \cdot & R(x^{-1}) \big) \\
 & = \Omega(x) \cdot \overline G(x).
\end{split}
\end{equation*}
With~\refeq{eq_list1_reverse_1}, we obtain:
\begin{equation*} 
 x^{n-k} \LPOL{0} + \LPOL{1} \cdot \overline R(x) = \Omega(x) \cdot \overline G(x),
\end{equation*}
which we can consider modulo $x^{n-k}$. We obtain
\begin{equation}\label{eq_KEcompl}
\LPOL{1} \cdot \overline R(x) \equiv \Omega(x) \cdot \overline G(x) \bmod x^{n-k}.
\end{equation}
Since $\overline G(0)\neq 0$, we can define the formal power series $\overline R(x)
/ \overline G(x)$:
\begin{equation}
\frac{\overline{R}(x)}{\overline{G}(x)}=\sum_{i=0}^\infty S_i x^i=S(x).
\end{equation}
Using the column multipliers~\refeq{eq_columnGRS} for the dual code, it
can be verified that $S(x)$ is the series of syndromes with
\begin{equation}
S_i=\sum_{j=1}^n \upsilon_j r_j \alpha_j^i.
\end{equation}
Thus, dividing~\refeq{eq_KEcompl} by $\overline G(x)$, we obtain
\begin{equation} \label{eq_KEcompl2}
 S(x) \cdot \LPOL{1} \equiv \Omega(x) \mod x^{n-k},
\end{equation}
which corresponds to the classical Key Equation~\refeq{eq_KeyEquation_frac}. The syndrome polynomial is $S(x)
\bmod \, x^{n-k}$, and $\LPOL{1}$ is the error-locator polynomial $\Lambda(x)$.

In the case of $\tau$ errors, we consider only the terms of the Key Equation of degree greater
than $n-k-\tau$ and we get the following homogeneous linear system of equations:
\begin{equation} \label{eq_classicalKE_matrix}
 \left(\begin{array}{cccc}
\mat[0]{S} & \mat[1]{S} & \dots & \mat[\tau]{S}\\
\mat[1]{S} & \mat[2]{S} & \dots & \mat[\tau+1]{S}\\
\vdots & \vdots & \iddots & \vdots \\
\mat[\tau-1]{S} & \mat[\tau]{S} & \dots & \mat[2\tau-1]{S}\\
\end{array}\right) 
\left(\begin{array}{c}
 {\LCOEFF{1}{\tau}} \\
 {\LCOEFF{1}{\tau-1}} \\
 \vdots \\ 
 {\LCOEFF{1}{0}} \\
\end{array}\right) = \mathbf{0}.
\end{equation}
The above syndrome matrix $\mat{S} = \vert \vert \mat[i,j]{S} \vert \vert $ for all $i
 \in \SETzero{\tau}$ and $j \in \SETzero{\tau+1} $ has Hankel form (see Definition~\ref{def_Hankel}).
Equation~\refeq{eq_KEcompl2} can be solved by the well-known Berlekamp--Massey algorithm~\cite{Berlekamp:AGT1968, Massey_Shift-registersynthesisandBCHdecoding_1969} 
or with a modification of the Extended Euclidean algorithm~\cite{Sugiyama_AMethodOfSolving_1975}. 
The parallels of the Berlekamp--Massey algorithm and the Extended Euclidean algorithm have
been considered in~\cite{Dornstetter_OnTheEquivalence_1987,Heydtmann_BMEuclidFIA_2000,Amoros-Sullivan:AAECC2009}.

We consider in the following the FIA~\cite{Feng_Tzeng1991}, that can be used to find the first $\mu+1$ linearly dependent columns
and connection coefficients $T_1, T_2, \dots, T_\mu$ for an arbitrary matrix.
The FIA allows a significant reduction of complexity when adjusted
to a Hankel matrix as in~\refeq{eq_classicalKE_matrix}. 

\subsection{The FIA for One Hankel Matrix}

\begin{figure*}[htb]
\centerline{\subfigure[Classic FIA (unadjusted)]{\scalebox{0.7}{\includegraphics{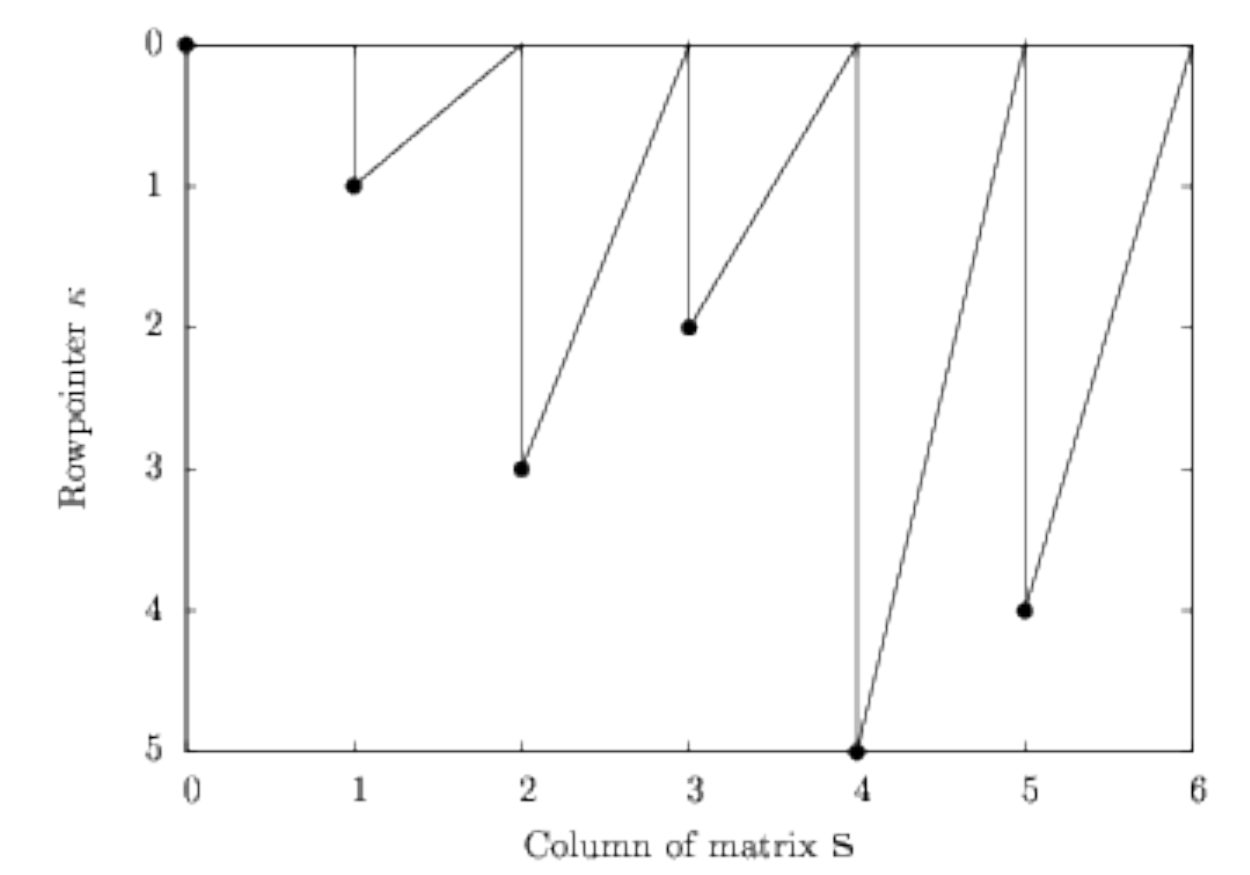}}
\label{fig_FIAHankel_unadjusted}}
\hfill
\subfigure[Adjusted FIA]{\scalebox{0.7}{\includegraphics{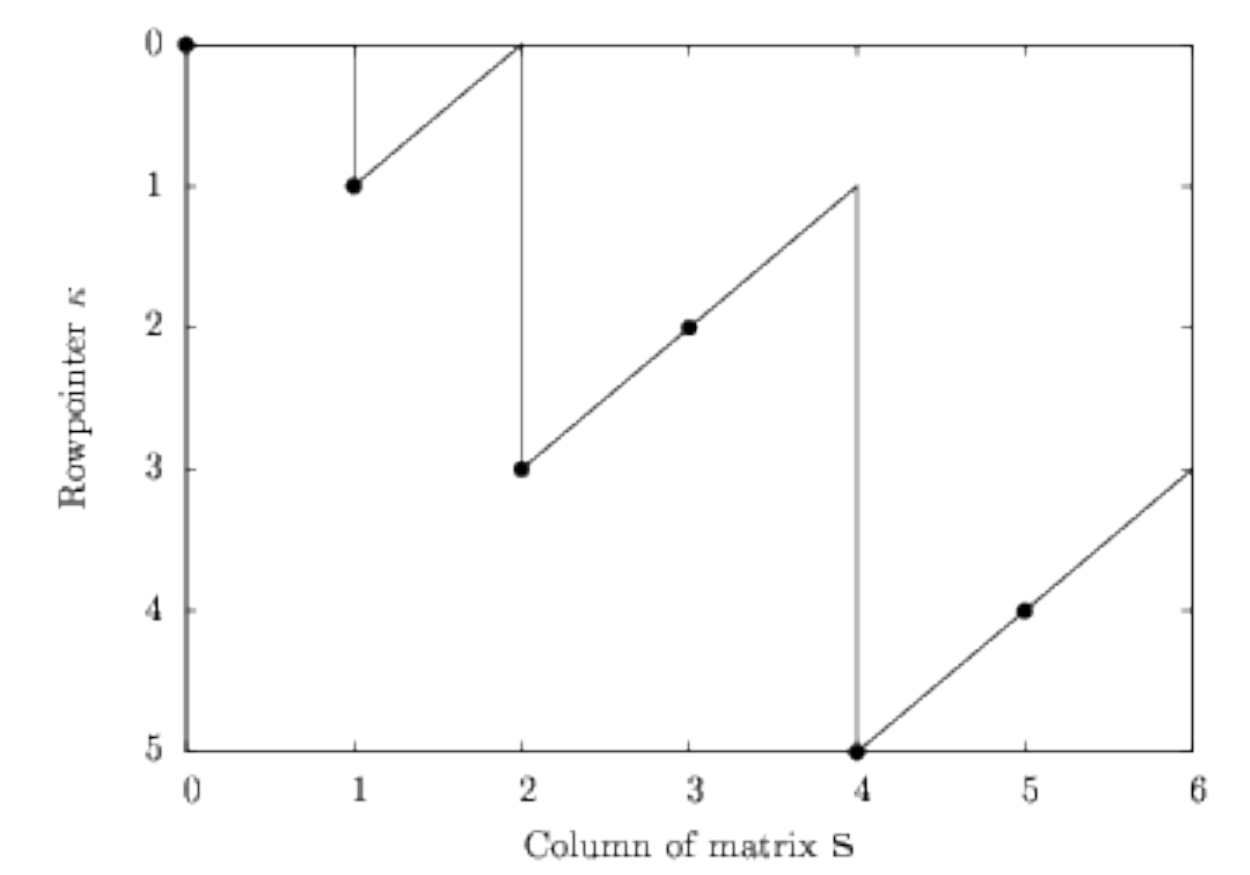}}
\label{fig_FIAHankel_adjusted}}}
\caption{Illustration of the row pointer $\kappa$ of the classic FIA
  (Sub-figure~\ref{fig_FIAHankel_unadjusted}) and of the adjusted
  FIA (Sub-figure~\ref{fig_FIAHankel_adjusted}) when both algorithms are
  applied to the same $6 \times 7$ Hankel syndrome matrix of a
  $\mathcal{GRS}(16,4)$ code. The dots indicate a true
  discrepancy. In this case, both algorithms enter a new column, but
  with different initial values of their row pointers.}
\label{fig_FIAHankel}
\end{figure*}

Given an arbitrary $m \times n$ matrix $\mat{A} = \vert \vert 
A_{i,j} \vert \vert$, the FIA outputs the minimal number of $\mu+1$
linearly dependent columns together with the polynomial $T(x) = \sum_{j=0}^{\mu} T_j
x^j$, with $T_\mu\neq0$, such that $ \sum_{j=0}^{\mu} T_j A_{i,j} = 0, \, i \in \SETzero{m} $ holds. The FIA scans the $\mu$th
column of the matrix $\mat{A}$ row-wise in the order $A_{0,\mu},
A_{1,\mu}, \dots$ and uses previously stored polynomials to update
the current polynomial $T(x)$. Let $\mu$ be the index of the current column
under inspection, and let $T(x) = \sum_{j=0}^{\mu} T_j
x^j$ be the current candidate polynomial that satisfies:
\begin{equation*}
  \sum_{j=0}^{\mu} T_j A_{i,j} = 0, \quad i \in \SETzero{\kappa}, 
\end{equation*}
for some value of the row index $\kappa$. In other words, the
coefficients of the polynomial $T(x)$ give us the vanishing
linear combination of the matrix consisting of the first $\kappa$
rows and the first $\mu+1$ columns of the matrix $\mat{A}$. Suppose
that the discrepancy:
\begin{equation} \label{eq_FIA_disc}
 \Delta = \sum_{j=0}^{\mu} T_j A_{\kappa,j}\neq0.
\end{equation}
for next row $\kappa$ is nonzero. 
If there exists a previously stored polynomial $T^{(\kappa)}(x)$ and a
nonzero discrepancy $\Delta^{(\kappa)}$, corresponding to row
$\kappa$, then the current polynomial $T(x)$ is updated in
the following way:
\begin{equation} \label{eq_FIA_update}
 T(x) \leftarrow T(x) - \frac{\Delta}{\Delta^{(\kappa)}} T^{(\kappa)}(x).
\end{equation}
The proof of the above update rule is straightforward~\cite{Feng_Tzeng1991}.

In the case $\Delta \neq 0$ and there is no discrepancy $\Delta^{(\kappa)}$ 
stored, the actual discrepancy $\Delta$ is stored as $\Delta^{(\kappa)}$.
The corresponding auxiliary polynomial is stored as $T^{(\kappa)}(x)$. Then, 
the FIA examines a new column $\mu+1$.

\begin{definition}[True Discrepancy]
Let the FIA examine the $\kappa$th row of the $\mu$th column of matrix \mat{A}.
Furthermore, let the calculated discrepancy~\refeq{eq_FIA_disc} be nonzero
and no other nonzero discrepancy be stored for row $\kappa$.
Then, the FIA examines a new column $\mu+1$. We call this case a \textbf{true} discrepancy.
\end{definition}

\begin{theorem}[Correctness and Complexity of the FIA~\cite{Feng_Tzeng1991}] \label{theo_correctcompl_FIA}
For an $m \times n $ matrix  with $n > m$, the Fundamental Iterative
Algorithm stops, when the row pointer has reached the last row of column $\mu$. Then,
the last polynomial $T^{(\mu)}(x)$ corresponds to a valid
combination of the first $\mu+1$ columns. The complexity of the
algorithm is $O(m^3)$.
\end{theorem}

For a Hankel matrix $\mat{S}$ (as in Definition~\ref{def_Hankel}), 
the FIA can be adjusted. Assume the case of a true discrepancy,
when the FIA examines the $\kappa$th row of the $\mu$th column of the structured matrix $\mat{S}$. The current
polynomial is $T(x)$.
Then, the FIA starts examining the $(\mu+1)$th column at row $\kappa-1$ with $T(x) \leftarrow x \cdot T(x)$ and not
at row zero. This reduces the cubic time complexity into a quadratic time complexity~\cite{Feng_Tzeng1991}.

To illustrate the complexity reduction of the FIA when adjusted to a
Hankel matrix (compared to the original, unadjusted FIA), we traced the examined rows 
for each column in Figure~\ref{fig_FIAHankel}. Figure~\ref{fig_FIAHankel_unadjusted} shows the values of
$\kappa$ of the FIA without any adaption. The row pointer $\kappa$ of the
adapted FIA is traced in Figure~\ref{fig_FIAHankel_adjusted}. 

The points on the lines in both figures indicate the case, where a true discrepancy has been
encountered.


\section{Sudan Interpolation Step with a Horizontal Band of Hankel Matrices} \label{sec_horizontalline}
\subsection{Univariate Reformulation of the Sudan Interpolation Step}
In this section, we recall parts of the work of Roth and Ruckenstein~\cite{Roth_Ruckenstein_2000, Ruckenstein_PHD2001}
for the interpolation step of the Sudan~\cite{sudan97decoding} principle. The aimed decoding radius is denoted by $\tau$, the corresponding list size is
$\listl$. 
\begin{problem}[Sudan Interpolation Step~\cite{sudan97decoding}] \label{prob:Sudan:interpol}
Let the aimed decoding radius $\tau$ and the received word $(r_1, r_2, \dots, r_n)$ be given.
The Sudan interpolation step determines a polynomial $Q(x,y)=\sum_{t=0}^{\listl} \QPOL{t}y^t \in\Fxy$, such that
\begin{enumerate}
\item $Q(x,y)\neq 0$,
\item \label{cond_Sudan_mult} $Q(\alpha_i,r_i/\upsilon^\prime_i) = 0,  \quad  \forall i \in \SET{n}$,
\item \label{cond_Sudan_wdeg} $\wdeg{1}{k-1}Q(x,y) < n-\tau $.
\end{enumerate}
\end{problem}
We present here a slightly modified version
of~\cite{Roth_Ruckenstein_2000}, to get an
appropriate basis for the extension to the interpolation step in the Guruswami--Sudan case.

\begin{figure*}[!b]
\setcounter{mytempeqncnt}{\value{equation}}
\normalsize
\setcounter{equation}{24}
\hrulefill
\begin{equation} \label{eq_SudanMatrix}
\mat{S}^{\prime} = \left( \begin{array}{cccc|ccccc|c|ccc}
{S}^{(0)}_{0} & {S}^{(0)}_{1} & \cdots & {S}^{(0)}_{k-2} & {S}^{(0)}_{k-1} & {S}^{(1)}_{0} & \cdots & {S}^{(0)}_{2k-3} & {S}^{(1)}_{k-2} & \cdots & \cdots & {S}^{(\listl-1)}_{N_{\listl-1}-1} & {S}^{(\listl)}_{N_{\listl}-1} \\
{S}^{(0)}_{1} & {S}^{(0)}_{2} & \cdots & {S}^{(0)}_{k-1} & {S}^{(0)}_{k} & {S}^{(1)}_{1} & \cdots & {S}^{(0)}_{2(k-2)} & {S}^{(1)}_{k-1} & \cdots & \cdots & {S}^{(\listl-1)}_{N_{\listl-1}} & {S}^{(\listl)}_{N_{\listl}} \\
\vdots & \vdots & \iddots & \vdots & \vdots & \vdots & \iddots & \vdots & \vdots & \iddots & \cdots & \vdots & \vdots\\
{S}^{(0)}_{n-1} & {S}^{(0)}_{n} & \cdots & {S}^{(0)}_{n + k-3} & {S}^{(0)}_{n - k-2} & {S}^{(1)}_{n-1} & \cdots & {S}^{(0)}_{n + 2(k-2)} & {S}^{(1)}_{n - k-3} & \cdots & \cdots &{S}^{(\listl-1)}_{n + N_{\listl-1}-2} & {S}^{(\listl)}_{n + N_{\listl}-2} \\
\end{array}\right)
\end{equation}
\setcounter{equation}{\value{mytempeqncnt}}
\end{figure*}

We have $\deg \QPOL{t} < N_t \defeq n-\tau-t(k-1), \; t \in \SETzero{\listl+1}$. Let $R(x)$ be the Lagrange
interpolation polynomial, s.t. $R(\alpha_i)=r_i/\upsilon^\prime_i$, $i
\in \SET{n}$ and $G(x)=\prod_{i=1}^n(x-\alpha_i)$. The reciprocal polynomial of $\QPOL{t}$ is denoted
by $\LPOL{t} = x^{N_t-1} Q^{(t)}(x^{-1})$.

Similar to Proposition~\ref{prop_WelchBerlekamp}, Roth--Ruckenstein~\cite{Roth_Ruckenstein_2000} proved
the following. There is an interpolation polynomial $Q(x,y)$ satisfying Conditions~\ref{cond_Sudan_mult} and~\ref{cond_Sudan_wdeg}
if and only if there exists a univariate polynomial $B(x)$ with degree smaller than $\listl(n-k) - \tau$, s.t. $Q(x,R(x)) = B(x)\cdot G(x)$.

Let the reciprocal polynomials be defined as in \refeq{eq_list1_reverse_1}.
From~\cite[Equation~(19)]{Roth_Ruckenstein_2000} we have:
\begin{equation} \label{eq_RRwithQ0}
\begin{split}
 \sum_{t=0}^\listl \LPOL{t} & \cdot x^{(\listl-t)(n-k)}  \cdot
 \overline R(x)^t \\ 
 & \equiv \overline{B}(x) \cdot \overline G(x) \bmod \, x^{n-\tau + \listl(n-k)},
\end{split}
\end{equation}
where $\deg \overline{B}(x) < \listl(n-k)-\tau$. We introduce the
power series
\begin{equation} \label{eq_powerseriesSudan}
T^{(t)}(x)\defeq\frac{\overline R(x)^t}{\overline
  G(x)}=\sum_{i=0}^\infty T^{(t)}_ix^i.
\end{equation}
Inserting~\refeq{eq_powerseriesSudan} into~\refeq{eq_RRwithQ0} leads to: 
\begin{equation} \label{eq_RRwithQ0Series}
\begin{split}
 \sum_{t=0}^\listl \LPOL{t} \cdot x^{(\listl-t)(n-k)} & \cdot
 T^{(t)}(x) \\
& \equiv \overline{B}(x) \bmod  x^{n-\tau + \listl(n-k)}.
\end{split}
\end{equation}
Based on~\refeq{eq_RRwithQ0Series} we can now define syndromes for 
Problem~\ref{prob:Sudan:interpol}.

\begin{definition}[Syndromes for Sudan] \label{def_syndRRwithQ0} 
The $\listl+1$ generalized syndrome polynomials $\SYNDPOL{t} \defeq \sum_{i=0}^{n+N_t-1}
S^{(t)}_i x^i$ are given by:
\begin{equation}
S_i^{(t)} = T^{(t)}_{i+(t-1)(n-1)}, \quad i\in \SETzero{n+N_t},\quad
t \in \SETzero{\listl+1}.
\end{equation}
The first order \textbf{Extended Key Equation} is: 
\begin{equation}
 \sum_{t=0}^\listl \LPOL{t} \cdot S^{(t)}(x)\equiv \overline{B}(x) \bmod
 \, x^{n-\tau + \listl(n-k)},
\end{equation}
with $\deg \overline{B}(x) < \listl(n-k)-\tau$.
\end{definition}
An explicit form of $S_i^{(t)}$ is:
\begin{equation} \label{eq_sudansyndexp}
\SYNDCOEFF{t}{i}=\sum_{j=1}  ^n \upsilon_j\frac{r_j^t}{{\upsilon'_i}^{t-1}}\alpha_j ^i, \quad i\in \SETzero{n+N_t},\quad
t \in \SETzero{\listl+1}.
\end{equation}

\begin{note}\label{note:RRDiagonal}
In~\cite{Roth_Ruckenstein_2000}, a further degree
reduction is proposed. Then~\refeq{eq_RRwithQ0Series}, is modulo $x^{n-k}$ and the polynomial \LPOL{0} disappears. 
We do not present this improvement here, because we cannot properly reproduce this behavior in the
Guruswami--Sudan case (see Note~\ref{note:GSDiagonal}).
\end{note}
The degree of the LHS of~\refeq{eq_RRwithQ0} is smaller
than $n-\tau + \listl(n-k)$. If we consider the terms of degree
higher than $ \listl(n-k)-\tau$, we obtain $n$ homogeneous linear
equations. Reverting back to the originals univariate polynomials $Q^{(t)}(x)$, we get the following system:
\begin{equation} \label{eq_RRwithQ0_explicit}
 \sum_{t=0}^\listl \sum_{i=0}^{N_t-1} Q^{(t)}_i \cdot S^{(t)}_{i+j} = 0, \quad j \in \SETzero{n}.
\end{equation}  
With $\mat{Q}^{(t)} =(Q^{(t)}_0,Q^{(t)}_1,\dots,Q^{(t)}_{N_t-1})^T$, we obtain the following matrix form:
\begin{equation} \label{eq_RRwithQ0matrix}
 \left( \mat{S}^{(0)} \, \mat{S}^{(1)} \, \cdots \, \mat{S}^{(\listl)} \right) \cdot \left(\begin{array}{c}
 \mat{Q}^{(0)} \\
 \mat{Q}^{(1)} \\
 \vdots \\ 
 \mat{Q}^{(\listl)} \\
\end{array}\right) = \mathbf{0},
\end{equation}
where each sub-matrix $\mat{S}^{(t)} = \vert \vert S^{(t)}_{i,j} \vert \vert,
\; i \in \SETzero{n}, j \in \SETzero{N_t}, t \in \SETzero{\listl+1}$
is a Hankel matrix. The $\listl+1$ syndrome polynomials $S^{(t)}(x) = \sum_{i=0}^{N_t-1}
S_i^{(t)}x^i$ of Definition~\ref{def_syndRRwithQ0} are associated with this horizontal band of $\listl+1$
Hankel matrices by $S^{(t)}_{i,j}=S^{(t)}_{i+j}$. 

In the following, we describe how the FIA can be adapted to solve
the homogeneous system of equations~\refeq{eq_RRwithQ0matrix}.

\subsection{Adjustment of the FIA for the Reformulated Sudan Interpolation Problem}
The FIA can directly be applied to the matrix $\vert \vert \mat{S}^{(0)} \, \mat{S}^{(1)} \, \cdots \, \mat{S}^{(\listl)} \vert \vert $ 
of~\refeq{eq_RRwithQ0matrix}, but if we want to take advantage of the
Hankel structure we have to scan the columns of $ \vert \vert \mat{S}^{(0)} \, \mat{S}^{(1)} \, \cdots \, \mat{S}^{(\listl)} \vert \vert $ 
in a manner given by the weighted degree requirement of the interpolation problem.

Let $\prec_H$ denote the ordering for the pairs $\{ (\nu,\mu) |
\nu \in \SETzero{\listl+1} \,\text{and} \, \mu \in \N \}$, where
$(\nu,\mu) \prec_H (\overline \nu, \overline \mu)$ is given by:
\begin{equation}\label{eq_prec_H}
\begin{split}
(\nu,\mu) & \prec_H(\overline \nu,\overline \mu) \iff  \\
& \left\{ \begin{array}{l}
 \nu+\mu(k-1) < \overline \nu + \overline \mu(k-1) \\
 \mbox{or} \\
 \nu + \mu(k-1) = \overline \nu + \overline \mu(k-1) \mbox{ and } \mu < \overline \mu.
 \end{array} \right.
\end{split}
\end{equation}
\setcounter{mytempeqncntstore}{\value{equation}}
\stepcounter{mytempeqncntstore}
\setcounter{equation}{\value{mytempeqncntstore}}

The pair that immediately follows $(\nu,\mu)$ with respect to the order
defined by $\prec_H$ is denoted by $\nexth{\nu}{\mu}$. The columns of the
matrix $\mat{S} = \vert \vert \mat{S}^{(0)} \, \mat{S}^{(1)} \,  \cdots \,  \mat{S}^{(\listl)} \vert \vert$ are reordered according to $\prec_H$. The pair
$(\nu,\mu)$ indexes the $\mu$th column of $\nu$th sub-matrix
$\mat{S}^{(\nu)}$. More explicitly, we obtain the following matrix
$\mat{S}^{\prime}$, where the columns of $\mat{S}$ are reordered (see Equation~\refeq{eq_SudanMatrix}).

The corresponding homogeneous system of equations can now be written in terms of
the inner product for bivariate polynomials (see Definition~\ref{def_InnerProduct}). 

\begin{problem}[Reformulated Sudan Interpolation Problem] \label{def_SystemHankelHorizontalRearranged}
Let the $\listl+1$ syndrome polynomials $\SYNDPOL{0}$,$\SYNDPOL{1}$,$\dots$, $\SYNDPOL{\listl}$ be
given by Definition~\ref{def_syndRRwithQ0} and let $S(x,y) \defeq \sum_{t=0}^{\listl} S^{(t)}(x)y^t$ be the corresponding bivariate
syndrome polynomial. We search a nonzero bivariate polynomial $T(x,y)$ such that:
\begin{equation}
\IP{x^{\kappa} T(x,y)}{S(x,y)} = 0,  \quad \kappa \in \SETzero{n}.
\end{equation}
\end{problem}
Hence, the bivariate polynomial $T(x,y)$ is a valid interpolation
polynomial for Problem~\ref{prob:Sudan:interpol}. Note that each polynomial $S^{(t)}(x)$, as defined
in~\refeq{eq_RRwithQ0}, has degree smaller than $N_t+n-1$. To index
the columns of the rearranged matrix $\mat{S}^{\prime}$, let 
\begin{equation} \label{eq_columncounter}
C_{\nu,\mu} = \big| \{ (t,i) \, | \, (t,i) \prec_H (\nu,\mu) \} \big|.
\end{equation}
Algorithm~\ref{FIA_RR} is the modified FIA for solving
Problem~\ref{def_SystemHankelHorizontalRearranged}. In contrast to the
original Roth--Ruckenstein adaption we consider all $n$ homogeneous
linear equations (instead of $\tau$), according to Note~\ref{note:RRDiagonal}. The
column pointer is given by $(\nu,\mu)$, for indexing the $\mu$th
column of the $\nu$th sub-matrix $\mat{S}^{(\nu)}$. Algorithm~\ref{FIA_RR}
virtually scans the rearranged matrix $\mat{S}^{\prime}$ column after
column (see Line~\ref{alg_SKE_calc_succ} of Algorithm~\ref{FIA_RR}).
The true discrepancy value for row $\kappa$ is stored in array $D$ as $D[\kappa]$, and the corresponding
intermediate bivariate polynomial is stored in array $A$ as
$A[\kappa]$. The discrepancy calculation and the update rule (see~\refeq{eq_FIA_disc} and~\refeq{eq_FIA_update} for the 
basic FIA) is adapted to the bivariate case (see Line~\ref{alg_SKE_update} of Algorithm~\ref{FIA_RR}). For
each sub-matrix $\mat{S}^{(\nu)}$, the previous value of the
row pointer $\kappa$ is stored in an array $R$ as
$R[\nu]$. We prove the initialization rule for the FIA solving
Problem~\ref{def_SystemHankelHorizontalRearranged} in the following
proposition.

\printalgo{
\SetVline
\linesnumbered
\SetKwInput{KwData}{\underline{Data structures}}
\SetKwInput{KwResult}{\underline{Initialize}}
\SetKwInput{KwIn}{\underline{Input}}
\SetKwInput{KwOut}{\underline{Output}}
\KwIn{Bivariate polynomial $S(x,y) = \sum_{t=0}^{\listl}S^{(t)}(x)y^t$;}
\KwOut{Bivariate polynomial $T(x,y)$;}
\BlankLine
\KwData{\\
Bivariate polynomial $T(x,y)= \sum_{t=0}^{\listl} T^{(t)}(x)y^{t}$;\\ 
Column pointers $(\nu,\mu)$, where $ \nu \in \SETzero{\listl}$, $\mu \in \SETzero{N_{\nu}}$;\\
Row pointer $\kappa \in \SETzero{n} $;\\
Array $D$ of $n$ entries in \F;\\
Array $R$ of $\listl+1$ entries in \N;\\
Array $A$ of $n$ entries in $\Fxy$;\\
Variable $\Delta \in \F$, variable \textit{compute} $\in \lbrace \mathrm{TRUE, FALSE} \rbrace$;
}
\BlankLine
\KwResult{\\
\For{$i \in \SETzero{n}$}{$D[i] \leftarrow 0$;}\\
\For{$i \in \SETzero{\listl+1}$}{$R[i] \leftarrow 0$;}\\
$(\nu,\mu) \leftarrow (0,0) $, $\kappa \leftarrow 0 $, 
\textit{compute} $ \leftarrow \mathrm{FALSE}$;
}
\BlankLine
\While{$ \kappa < n $}{
\eIf{\textit{compute} }{$\Delta \leftarrow \langle x^{\kappa}\cdot T(x,y), S(x,y) \rangle $\;}
{
  \eIf{$R[\nu] < 1$}
  {
  $ T(x,y) \leftarrow y^{\nu} \cdot x^{\mu} $\;
  $ \Delta \leftarrow S^{(\nu)}_{\mu}$\;
  $ \kappa \leftarrow 0$\;
  }
  {
  $ T(x,y) \leftarrow x \cdot A[R[\nu]](x,y)$\;\nllabel{alg_SKE_jump1}
  $ \Delta \leftarrow D[R[\nu]]$\; \nllabel{alg_SKE_jump2}
  $ \kappa \leftarrow R[\nu]-1$\; \nllabel{alg_SKE_jump3}
  }
  $\text{\textit{compute}} \leftarrow \mathrm{TRUE}$\;
}
\eIf {$\Delta = 0$ or $D[\kappa] \neq 0$}
{ 
 \If{$\Delta \neq 0 $}{$T(x,y) \leftarrow T(x,y) - \frac{\Delta}{D[\kappa]} \cdot A[\kappa](x,y)$\; \nllabel{alg_SKE_update} }
 $\kappa \leftarrow \kappa +1 $\; 
}(\textit{ /* $\Delta \neq 0$ and $D[\kappa] = 0$  */})
{
$ A[\kappa] \leftarrow T(x,y)  $\;
$ D[\kappa] \leftarrow \Delta$\;
$ R[\nu] \leftarrow \kappa$\; \nllabel{alg_SKE_store_row}
$ \text{\textit{compute}} \leftarrow \mathrm{FALSE}$\;
$ (\nu,\mu) \leftarrow \nexth{\nu}{\mu} $\; \nllabel{alg_SKE_calc_succ}
}
}
\caption{Algorithm Solving Problem~\ref{def_SystemHankelHorizontalRearranged}}
\label{FIA_RR}
}

\begin{figure*}[htb] \centering
   \begin{minipage}{0.5\textwidth} 
   \centering 
       \includegraphics[width = \columnwidth]{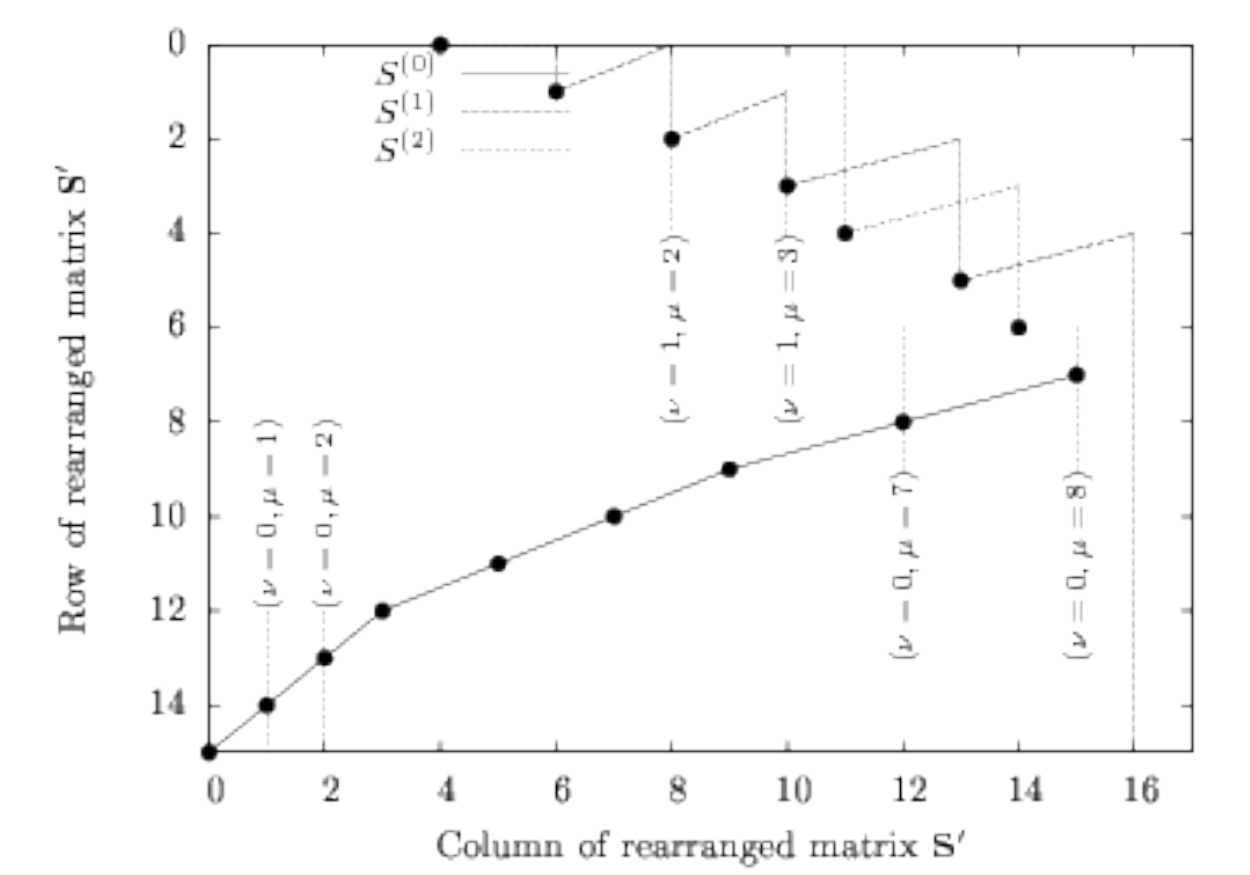}
        \caption{Illustration of the row pointer $\kappa$ of
		Algorithm~\ref{FIA_RR} applied to a horizontal band of three
		Hankel matrices $\mat{S}^{(0)}$, $\mat{S}^{(1)}$ and
		$\mat{S}^{(2)}$. The columns of the $16 \times 18$ matrix
		$\mat{S}^{\prime}$ are arranged under $\prec_H$-ordering. The
		three lines $S^{(0)}$, $S^{(1)}$ and $S^{(2)}$ trace the row
		pointer for each sub-matrix $\mat{S}^{(0)}$, $\mat{S}^{(1)}$
		and $\mat{S}^{(2)}$.}      
        \label{fig_ExampleHorizont}
    \end{minipage}
\hfill
   \begin{minipage}{0.45\textwidth} 
      \renewcommand{\arraystretch}{1.2}
      \centering
      \captionof{table}{Column-index $C_{\nu,\mu}$ and column pointer $(\nu,\mu)$ of the re-arranged matrix $\mathbf{S}^{\prime}$ of the reformulated Sudan 
      interpolation step for a $\mathcal{GRS}(16,4)$ code with decoding radius $\tau=7$ and list size $\listl=2$.} 
      \begin{tabular}{cccc}
      \hline
      \multicolumn{4}{c}{\textbf{Column $C_{\nu,\mu}$ and}}  \\
      \multicolumn{4}{c}{\textbf{Column pointers $(\nu,\mu)$}}  \\
      \hline\hline
0 & (0,0) & 9  & (0,6) \\
1 & (0,1) & 10 & (1,3) \\
2 & (0,2) & 11 & (2,0) \\
3 & (0,3) & 12 & (0,7) \\
4 & (1,0) & 13 & (1,4) \\
5 & (0,4) & 14 & (2,1) \\
6 & (1,1) & 15 & (0,8) \\ 
7 & (0,5) & 16 & (1,5) \\
8 & (1,2) & 17 & (2,2) \\
      \hline
      \end{tabular}
      \label{tb_Ex_Horizontal}
    \end{minipage}
\end{figure*}

\begin{proposition}[Initialization Rule] \label{prop_initSKE}
Assume Algorithm~\ref{FIA_RR} examines column $(\nu,\mu)$ of a
syndrome matrix $\mat{S} = \vert \vert \mat{S}^{(0)} \, \mat{S}^{(1)} \, \cdots
\, \mat{S}^{(\listl)} \vert \vert $ as defined in~\refeq{eq_RRwithQ0matrix} (or equivalently the bivariate polynomial
$S(x,y)$). Assume that a true discrepancy is obtained in row
$\kappa_{\nu}$. 

Let $(\overline{\nu},\overline \mu) = \nexth{\nu}{\mu}$. 
Hence, Algorithm~\ref{FIA_RR} can examine column $(\overline \nu,\overline \mu)$ at row $\kappa_{\overline \nu}-1$ with the
initial value $\overline{T}(x,y) = x \cdot T(x,y) $ , where $\kappa_{\overline \nu}$ is the
index of the row, where the last true discrepancy in the $\overline \nu$th sub-matrix $\mat{S}^{(\overline \nu)}$ 
was calculated. The polynomial $T(x,y)$ is the stored intermediate polynomial for $\mat{S}^{(\overline \nu)}$, i.e., $\kappa_{\overline \nu}=R[\overline \nu] $ and
$T(x,y)=A[\kappa_{\overline \nu}]$.
\end{proposition}
\begin{IEEEproof} \label{proof_FIA_RRHankel}
In terms of the inner product (see Definition~\ref{def_InnerProduct}), we have:
\begin{equation*}
\IP{x^{i} T(x,y)}{S(x,y)} = 0, \quad i \in \SETzero{\kappa_{\overline \nu}}.
\end{equation*}
Let us write $T(x,y)=\sum_{t=0}^{\overline \nu} \sum_{j=0}^{\overline \mu}  T^{(t)}_{j} x^j y^t$. We have
$\overline T (x,y)=\sum_{t=0}^{\overline \nu} \sum_{j=1}^{\overline \mu + 1 } T^{(t)}_{j-1} x^j y^t$ and we compute:
\begin{align*}
\IP{x^{i} \overline{T}(x,y)}{S(x,y)} & = \sum_{t=0}^{\overline \nu} \sum_{j=1}^{\overline \mu +1} \overline T^{(t)}_{j} \cdot S^{(t)}_{i,j} \\
& = \sum_{t=0}^{\overline \nu} \sum_{j=1}^{\overline \mu +1}  T^{(t)}_{j-1} \cdot S^{(t)}_{i,j} \\
& = \sum_{t=0}^{\overline \nu} \sum_{j=0}^{\overline \mu}  T^{(t)}_{j} \cdot S^{(t)}_{i+1,j} \quad \textit{(Hankel)} \\
& = \IP{x^{i+1} T(x,y)}{S(x,y)} 
\end{align*}
which is zero for the rows of index $i\in\SETzero {\kappa_{\overline \nu}-1}$.

\end{IEEEproof}
Similarly to the FIA for one Hankel matrix we can start examining a new $\overline \mu$th column of 
the sub-matrix $\mat{S}^{({\overline \nu})}$ in row $\kappa_{\overline \nu}-1$. Note that the
previous value of the row pointer $\kappa_{\overline \nu}$ is stored in $R[\overline \nu]$. 

Before Algorithm~\ref{FIA_RR} enters a new
column, the coefficients of the intermediate bivariate connection
polynomial $T(x,y)$ give us the vanishing linear combination of the
sub-matrix consisting of the first $\kappa_{\nu}$ rows and $C_{\nu, \mu}$
previous columns of the rearranged matrix $\mathbf{S}^{\prime}$ (see~\refeq{eq_SudanMatrix}).
The following theorem summarizes the properties of
Algorithm~\ref{FIA_RR}.

\begin{theorem}[Algorithm~\ref{FIA_RR}] 
Let $\mat{S} = \vert \vert \mat{S}^{(0)} \, \mat{S}^{(1)} \, \cdots \, \mat{S}^{(\listl)} \vert \vert $ 
be the $n \times \sum_{t=0}^{\listl} N_t$ matrix as defined in~\refeq{eq_RRwithQ0matrix} and $S(x,y)$ the
associated bivariate syndrome polynomial for the reformulated Sudan
interpolation problem. Algorithm~\ref{FIA_RR} returns a bivariate
polynomial $T(x,y)$ such that:
\begin{equation*}
\IP{x^\kappa T(x,y)}{S(x,y)} = 0, \quad \kappa \in \SETzero{n}.
\end{equation*}
The time complexity of Algorithm~\ref{FIA_RR} is $\ON{\listl n^2}$. 
\end{theorem}
\begin{IEEEproof}
The correctness of Algorithm~\ref{FIA_RR} follows from the correctness of the
basic FIA (see Theorem~\ref{theo_correctcompl_FIA}) and from the correctness of the initialization rule (Proposition~\ref{prop_initSKE}) 
considering that Algorithm~\ref{FIA_RR} deals with the column-permuted
version $\mat{S}^{\prime}$ of the original matrix $\mat{S} =  \vert \vert \mat{S}^{(0)} \, \mat{S}^{(1)} \, \cdots \, \mat{S}^{(\listl)} \vert \vert $.

The proof of the complexity of Algorithm~\ref{FIA_RR} is as follows. We trace the
triple:
\begin{equation*}
\big( (\nu,\mu),\left(\kappa_0,\kappa_1, \dots,\kappa_\listl\right),\delta \big).
\end{equation*}
where $(\nu,\mu)$ is the current column pointer of Algorithm~\ref{FIA_RR} examining
the $\mu$th column of the $\nu$th sub-matrix $\mat{S}^{(\nu)}$. The variables $\kappa_0$,$\kappa_1$,$\dots$,$\kappa_\listl$ are
the values of the last row reached in the sub-matrices $\mat S^{(0)},\mat S^{(1)},\dots,\mat S^{(\listl)}$.
These values are stored in the array $R$ in Algorithm~\ref{FIA_RR}.
The value $\delta$ is the number of already encountered true
discrepancies of Algorithm~\ref{FIA_RR}. 
Assume $(\nu,\mu)$ is the current column pointer of Algorithm~\ref{FIA_RR}.
The two following events in Algorithm~\ref{FIA_RR} can happen:\\
1) Either, there is no true discrepancy, then Algorithm~\ref{FIA_RR}
stays in the same column and $\kappa_\nu$ increases by one. The triple becomes
\begin{equation*}
\big((\nu,\mu),(\kappa_0,\kappa_1,\dots,\kappa_{\nu} \leftarrow \kappa_{\nu}+1,\dots,\kappa_\listl),\delta \big).
\end{equation*}
2) Or, there is a true discrepancy, then Algorithm~\ref{FIA_RR} examines
column $(\overline \nu,\overline \mu)=\nexth{\nu}{\mu}$ and the triple becomes
\begin{equation*}
\big((\overline \nu, \overline \mu),(\kappa_0,\kappa_1,\dots,\kappa_{\overline \nu} \leftarrow \kappa_{\overline \nu}-1,\dots,\kappa_\listl),\delta \leftarrow \delta+1 \big).
\end{equation*}
For both cases, the sum $\mathsf{Iter}$ over the triple is
\begin{equation}
\mathsf{Iter} = C_{\nu,\mu} + \left( \sum_{t \in \SETzero{\listl+1}} \kappa_t \right) + \delta,
\end{equation}
when Algorithm~\ref{FIA_RR} examines the $(\nu,\mu)$th column of the matrix $\vert \vert \mat{S}^{(0)} \, \mat{S}^{(1)} \, \cdots \, \mat{S}^{(\listl)} \vert \vert $.
From~\refeq{eq_columncounter}, we have $C_{\nexth{\nu}{\mu}} = C_{(\nu, \mu)} + 1$.
The sum $\mathsf{Iter}$ increases by one in each iteration of Algorithm~\ref{FIA_RR}. The initial value of $\mathsf{Iter}$
is zero and the last value can be bounded by: 
\begin{equation*}
\mathsf{Iter} <  \ON{n} + \ON{\listl n} + \ON{n} \leq \ON{\listl n}.
\end{equation*}
Each discrepancy computation costs $O(n)$ and Algorithm~\ref{FIA_RR} does not have to examine more than the $(n+1)$th
columns of the $n \times \sum_{t=0}^{\listl}N_t$ matrix $\vert \vert \mat{S}^{(0)} \, \mat{S}^{(1)} \, \cdots \, \mat{S}^{(\listl)} \vert \vert $. 
Thus, the total cost of Algorithm~\ref{FIA_RR} is $O(\listl n^2)$.
\end{IEEEproof}
In the following, we illustrate the values of the row pointer $\kappa$
of Algorithm~\ref{FIA_RR}, when applied to a syndrome matrix $\mat{S} = \vert \vert \mat{S}^{(0)} \, \mat{S}^{(1)} \, \mat{S}^{(2)} \vert \vert $ that consists of three Hankel matrices.

\subsection{Example: Sudan Decoding of a Generalized Reed--Solomon Code with Adapted FIA} \label{subsec_ex_SKE}
We consider a \GRS{16}{4} code over \GF{17}. 
For a decoding radius $\tau=7=\dhalf+1$, the list size is $\listl=2$. The degrees of the three univariate polynomials
$\QPOL{0}$, $\QPOL{1}$ and $\QPOL{2}$ are limited to $(N_0,N_1,N_2) =
(9,6,3)$ and we have more unknowns than interpolation constraints ($N_0+N_1+N_2 > n$).

Figure~\ref{fig_ExampleHorizont} illustrates the row pointer of Algorithm~\ref{FIA_RR} when the $16 \times 18$ syndrome matrix $\vert \vert  \mat{S}^{(0)} \, \mat{S}^{(1)} \, \mat{S}^{(2)}  \vert \vert $ is examined.
The columns of the syndrome matrix are virtually rearranged according to the $\prec_H$-ordering and
Algorithm~\ref{FIA_RR} scans the re-arranged matrix $\mat{S}^{\prime}$ column by column. The column-index $C_{\nu,\mu}$ (see~\refeq{eq_columncounter})
and the corresponding column pointer $(\nu,\mu)$ are listed in
Table~\ref{tb_Ex_Horizontal}.

The three zig--zag lines $S^{(0)}$, $ S^{(1)}$ and $S^{(2)}$ in Figure~\ref{fig_ExampleHorizont} trace the
value of the row pointer $\kappa$ for the three sub-matrices
$\mat{S}^{(0)}$, $\mat{S}^{(1)}$ and $\mat{S}^{(2)}$, which have a
Hankel structure.  The dots indicate the case, where a true
discrepancy occurs. After the $k$th column (here $k-1 = 3$),
every second column corresponds to the same sub-matrix.

After column $10$ of the rearranged matrix $\mat{S}^{\prime}$, every
third column of $\mat{S}^{\prime}$ corresponds to the same sub-matrix
$\mat{S}^{(\nu)}$. 
Let us investigate two cases, where a true discrepancy in Algorithm~\ref{FIA_RR} occurs.
They are marked in column $C_{0,7} = 12$ and $C_{0,8}=15$ 
of the re-arranged $\mat{S}^{\prime}$ in Figure~\ref{fig_ExampleHorizont}.
In between column 12 and 15 one column of the sub-matrices $\mat{S}^{(1)}$ and $\mat{S}^{(2)}$
is examined by Algorithm~\ref{FIA_RR}. In column $(0,8)$, Algorithm~\ref{FIA_RR}
starts investigating the second row, because the true discrepancy in column 
$(0,7)$ occurred in the third row (according to Proposition~\ref{prop_initSKE}).

\subsection{The FIA for a Vertical Band of Hankel Matrices} \label{sec_verticalline}
The FIA can also be adapted to a matrix consisting of Hankel matrices
arranged vertically. This case has been
considered for example in~\cite{Schmidt_CollaborativeDecoding_2009, Zeh_EfficientListDecoding_2009}.  
The basic idea for such a vertical band of Hankel matrices is the same as in the
previous case. The rows of each sub-matrix of Hankel structure are scanned in a similar
interleaving order as the columns of the previous case.  

The obtained time complexity for a vertical band of $s$ Hankel matrices, where
each sub-matrix consist of $N$ columns, is $\ON{sN^2}$.


\section{Guruswami--Sudan Interpolation Step with a Block-Hankel Matrix} \label{sec_GSAndHankelMatrices}

\subsection{The Guruswami--Sudan Interpolation Step for Generalized Reed--Solomon Codes}

We consider again a Generalized Reed--Solomon code with support set
$\alpha_1,\alpha_2, \dots,\alpha_n$, multipliers
$\upsilon^{\prime}_1,\upsilon^{\prime}_2, \dots,\upsilon^{\prime}_n$ and dimension $k$, as introduced in
Section~\ref{sec_defs}. Let $\upsilon_1,\upsilon_2, \dots,\upsilon_n$ according to~\refeq{eq_columnGRS} be the
multipliers of the dual Generalized Reed--Solomon code.

Let $(r_1,r_2,\dots,r_n)$ be the received word. The Guruswami--Sudan
decoding principle~\cite{GuruswamiSudan_ImproveddecodingofReed-Solomonandalgebraic-geometrycodes_1999,Guruswami_ListDecodingofError-CorrectingCodes_1999,Guruswami_ALGORITHMICRESULTSINLISTDECODING_2007} improves the previous algorithms by
introducing an additional parameter $s$, which is the order of multiplicity for the $n$ points $(\alpha_1, r_1/\upsilon^{\prime}_1), (\alpha_2, r_2/\upsilon^{\prime}_2), \dots,
(\alpha_n, r_n/\upsilon^{\prime}_n) $. The parameter $s$
influences the decoding radius $\tau$ and the list size
$\listl$. The relationship between these parameters has been discussed in many publications (see e.g.~\cite{McEliece_TheGuruswami-SudanDecodingAlgorithmforReed-SolomonCodes_2003}).
\begin{problem}[Guruswami--Sudan Interpolation Step~\cite{GuruswamiSudan_ImproveddecodingofReed-Solomonandalgebraic-geometrycodes_1999}]\label{prob:GS:interpol}
Let the aimed decoding radius $\tau$, the multiplicity $s$ and the received word $(r_1, r_2, \dots, r_n)$ be given. The Guruswami--Sudan interpolation step determines
a polynomial $Q(x,y)=\sum_{t=0}^{\listl} \QPOL{t}y^t \in\Fxy$, such that
\begin{enumerate}
\item $Q(x,y)\neq 0$,
\item\label{cond:GS:mult} $\mult{}{\alpha_i,r_i/\upsilon^{\prime}_i}\geq s, \quad \forall i \in \SET{n}$,
\item\label{cond:GS:wdeg} $\wdeg{1}{k-1} Q(x,y) < s(n-\tau) $.
\end{enumerate}
\end{problem}
As in the previous section, let $N_t$ denote the degree of the
$\listl+1$ univariate polynomials $\QPOL{t}$. From Condition~\ref{cond:GS:wdeg}) of Problem~\ref{prob:GS:interpol} we get:
\begin{equation} \label{eq_defNt}
\deg \QPOL{t} < N_t \defeq s(n-\tau)-t(k-1),\quad t\in\SETzero{\listl+1}.
\end{equation}

\subsection{Univariate Reformulation of the Guruswami--Sudan Interpolation Problem and A Block-Hankel Matrix} \label{sec_GSKE}
We reformulate the Guruswami--Sudan interpolation
problem to obtain not one, but a system of several Extended Key
Equations. The corresponding homogeneous linear system has a Block-Hankel form.

\begin{proposition}[Univariate Reformulation] \label{prop_propositionGS}
Let the integers $s$, $\tau$, $\listl$ and the received vector $(r_1,r_2,\dots,r_n)$ be given. Let $R(x)$ be the
Lagrange interpolation polynomial, such that
$R(\alpha_i)=r_i/\upsilon^\prime_i$, $i \in \SET{n}$. Let $G(x)=\prod_{i=1}^n(x-\alpha_i)$.
A polynomial $Q(x,y)$ satisfies Conditions~\ref{cond:GS:mult}) and~\ref{cond:GS:wdeg}) of
Problem~\ref{prob:GS:interpol}, if and only if there exist $s$
polynomials $B^{(b)}(x) \in \Fx$ such that
\begin{equation} \label{eq_GSKE}
Q^{[b]}(x,R(x)) = B^{(b)}(x) \cdot G(x)^{s-b},
\end{equation}
and $\deg B^{(b)}(x) < \listl (n-k) - s\tau+b$, $ b \in \SETzero{s}$.
\end{proposition}
Note that $Q^{[b]}(x,y)$ denotes the $b$th
Hasse derivative of the bivariate polynomial $Q(x,y)$ with respect to
the variable $y$ (see Definition~\ref{def_multiplicity}).

We first prove the following lemma.
\begin{lemma} \label{lem_reformulationGS_1}
Let $(\alpha_i,r_i) \in \F^{2}$ be given, and let $R(x)\in\Fx$ be any polynomial
such that $R(\alpha_i)=r_i$. A polynomial $Q(x,y)$ has
multiplicity at least $s$ at $(\alpha_i,r_i)$ if and only if
$(x-\alpha_i)^{s-b} | \UHASSE{b}(x,R(x))$, $\forall \, b\in \SETzero{s}$.
\end{lemma}
\begin{IEEEproof}
After translation to the origin, we can assume that
$(\alpha_i,r_i)=(0,0)$, and $R(0)=0$, i.e., $x|R(x)$. Let
$Q(x,y)=\sum_{i}Q_i(x,y)$, where $Q_i(x,y)$ is homogeneous of degree
$i$.

We first suppose that $Q(x,y)$ has at least a multiplicity $s$ at
$(0,0)$, i.e., $Q_i(x,y)=0$, for $i<s$. Hence, we have
\begin{equation*}
Q^{[b]}(x,R(x)) = \sum_{i\geq s-b} Q^{[b]}_i(x,R(x)).
\end{equation*}
For $b<s$, the polynomials $Q_i^{[b]}(x,y)$ have no terms of degree less than
$s-b$, and with $x|R(x)$, we have $x^{s-b}|Q^{[b]}_i(x,R(x))$.
It follows, that $x^{s-b}$ divides $Q^{[b]}(x,R(x))$ for all $ b\in \SETzero{s}$.

Suppose for the converse that $x^{s-b} | Q^{[b]}(x,R(x))$. That is,
$Q^{[b]}(x,R(x))=x^{s-b}U^{(b)}(x)$, for some polynomials $U^{(b)}(x)$ and
$b \in \SETzero{s}$. Using Taylor's formula with the Hasse
derivatives~\cite[p. 89]{Roth_IntroductiontoCodingTheory_2006} we have:
\begin{align*}
Q(x,y) = & \sum_bQ^{[b]}(x,R(x))\cdot(y-R(x))^b\\
       = & \sum_{b < s}Q^{[b]}(x,R(x))\cdot(y-R(x))^b \\
       & +\sum_{b \geq s}Q^{[b]}(x,R(x))\cdot(y-R(x))^b\\
      = & \sum_{b < s}x^{s-b}U^{(s- b)}(x)\cdot(y-R(x))^{b} \\
      &+\sum_{b \geq s}Q^{[b]}(x,R(x))\cdot(y-R(x))^b\label{eq:final:eq:div}.
\end{align*}
Now, $(y-R(x))^{b}$ has only terms of degree higher than $b$, since
$x|R(x)$. Thus, we have no terms of degree less than $s$
in $Q(x,y)$.
\end{IEEEproof}
\begin{IEEEproof}[Proof of Proposition~\ref{prop_propositionGS}]
From the previous lemma, we know that $(x-\alpha_i)^{ s-b} |
Q^{[b]}(x,R(x))$, $b\in\SETzero{s}$, $i\in \SET n$. Since all $(x-\alpha_i)$'s
are distinct the Chinese Remainder Theorem for univariate polynomials implies that $G(x)^{s-b}|Q^{[b]}(x,R(x))$. 
The degree condition follows easily.
\end{IEEEproof}

Proposition~\ref{prop_propositionGS} enables to
rewrite the $s$ equations~\refeq{eq_GSKE} more explicitly:
\begin{equation} \label{eq_Q_GSKE}
\begin{split}
 \sum_{t=b}^{\listl} \binom{t}{b} \QPOL{t} & \cdot R(x)^{t-b} = \\
& B^{(b)}(x) \cdot G(x)^{s-b},\quad  b \in \SETzero{s}.
\end{split}
\end{equation}
As usual, let the reciprocal polynomials be:
\begin{equation*} \label{eq_list1_reverse}
 \begin{split}
  \overline R(x) & = x^{n-1} \cdot R(x^{-1}), \\
  \overline G(x) & = x^n \cdot G(x^{-1}) = \prod_{i=1}^n(1-\alpha_ix), \\
  \overline B^{(b)}(x) & = x^{\listl(n-k)-s\tau-b-1} \cdot B(x^{-1}), \\
  \LPOL{t} & = x^{N_t-1} \cdot Q^{(t)}(x^{-1}).
 \end{split}
\end{equation*}
Inserting them into~\refeq{eq_Q_GSKE}, leads to:
\begin{equation} \label{eq_GSKE_Lamda}
\begin{split}
\sum_{t=b}^{\listl} \binom{t}{b}  \LPOL{t} & \cdot x^{(\listl-t)(n-k)} \cdot
\overline{R}(x)^{t-b} \\
& = \overline B^{(b)}(x) \cdot \overline{G}(x)^{s-b}, 
\end{split}
\end{equation}
Since $G(x)$ is relatively prime to $x^{(\listl-b)(n-k)}$, it admits
an inverse modulo $x^{(\listl-b)(n-k)}$. The Taylor series of
$\overline R(x)^{t-b}/ \overline G(x)^{s-b}$ is denoted by
$\SYNDPOLb{b,t}$. Then~\refeq{eq_GSKE_Lamda} leads to $s$ equations:
\begin{equation*} 
\begin{split}
\sum_{t=b+1}^{\listl} \binom{t}{b} \LPOL{t} & \cdot x^{(\listl-t)(n-k)}  \cdot T^{(b,t)}(x) \\
& \equiv \overline B^{(b)}(x) \bmod \, x^{(\listl-b)(n-k)}, \quad b \in \SETzero{s}.
\end{split}
\end{equation*}
where each equation is denoted by $\EKE{b}$.
Note that the degree of $\overline B^{(b)}(x)$ can be greater than $(\listl-b)(n-k)$ and it is not clear
how to properly truncate this identity, as in~\cite{Roth_Ruckenstein_2000, Ruckenstein_PHD2001}, noted in Note~\ref{note:RRDiagonal},
or as in the case of the classical Key Equation (see Section~\ref{sec_WBandFIA}). 

In the following, we consider the complete system of $\binom{s+1}{2}n$ homogeneous linear equations. We have $\deg Q^{[b]}(x,R(x)) = s(n-
\tau) + \listl(n-k)-b(n-1)$. We obtain $s$ equations for the $b$th derivative with the following truncation:
\begin{equation} \label{eq_GSKE_full}
\begin{split}
& \sum_{t=b}^{\listl}  \binom{t}{b}  \LPOL{t} \cdot  x^{(\listl-t)(n-k)} \cdot T^{(b,t)}(x) \\ 
& \equiv \overline B^{(b)}(x) \bmod \, x^{s(n-\tau)+\listl(n-k)-b(n-1)}, \quad b \in \SETzero{s}.
\end{split}
\end{equation}
Let us write $\EKEzero{b}$ for the $b$th equation as above.

\begin{proposition}
  Let $d=n-k+1$ be the minimum distance of the considered \GRS{n}{k} code. Let $b$ be such that $s\tau-bd\geq 0$.  If
  $\LPOL{b+1}, \dots,\LPOL{\listl}$ is a solution to $\EKE{b}$,
  then there exists $\LPOL b$ such that
  $\LPOL{b},\LPOL{b+1},\dots,\LPOL{\listl}$ is a solution to $\EKEzero{b}$.
\end{proposition}
\begin{IEEEproof} 
Let us consider~\refeq{eq_Q_GSKE}. We isolate $Q^{(b)}(x)$ and get
\begin{equation}
\begin{split}
Q^{(b)}(x)+\sum_{t=b+1}^{\listl} \binom{t}{b} Q^{(t)}(x) & \cdot R(x)^{t-b} \\
&=B^{(b)}(x) \cdot G(x)^{s-b}.
\end{split}
\end{equation}
and thus $Q^{(b)}(x)$ is the remainder of the Euclidean division of
$\sum_{t=b+1}^{\listl} \binom{t}{b} Q^{(t)}(x)R(x)^{t-b}$ by $G(x)^{s-b}$, as
long  as $\deg Q^{(b)}(x) < \deg G(x)^{s-b}$, which gives
$s(n-\tau)-b(k-1) \leq (s-b)n$, i.e., $s\tau-bd \geq 0$.
\end{IEEEproof}

\begin{note}\label{note:GSDiagonal}
We denote $b_0=\lfloor (s\tau)/d \rfloor$. Actually, we can
consider~\refeq{eq_GSKE_Lamda} and substitute the $\LPOL{b}$, for
$b\in\SET{b_0}$, successively. This is possible for the case of the
first order system ($s=1$), noted in Note~\ref{note:RRDiagonal}. In
the more general Guruswami--Sudan case, we can obtain a reduced system with $\LPOL{b_0+1},\dots,\LPOL{\listl}$, but it seems
that this reduced system lost its Block-Hankel structure.
Thus, there are no benefits of reducing the number of unknowns. 
We could not find a proper interpretation of the quantity $b_0=\lfloor (s\tau)/d \rfloor$.
\end{note}

With~\refeq{eq_GSKE_full}, we now can define the syndrome polynomials for the reformulated
Guruswami--Sudan interpolation problem.

\begin{definition}[Syndromes for Guruswami--Sudan] \label{def_syndGSwithQ0}
The syndrome polynomials $\SYNDPOL{0,0},\SYNDPOL{0,1},\dots,\SYNDPOL{0,\listl},\SYNDPOL{1,1},\dots,\SYNDPOL{s-1,\listl}$
with $\SYNDPOL{b,t}= \sum_{i=0}^{(s-b)n+N_t-1} S^{(b,t)}_i x^i$ are given by:
\begin{equation}
\begin{split}
S_i^{(b,t)} = T^{(b,t)}_{i+\left(b+1+t(n-1)-sn\right)},\quad & b \in \SETzero{s},\\
& t=b,\dots,\listl,
\end{split}
\end{equation}
where $\SYNDPOLb{b,t}$ denotes the power series of $\overline R(x)^{t-b}/ \overline G(x)^{s-b}$.

The ($s$th order) \textbf{Extended Key Equations} are
\begin{equation} \label{eq_GSKE_Synd}
\begin{split}
 \sum_{t=b}^\listl \LPOL{t} & \cdot S^{(b,t)}(x) \\
 & \equiv \overline{B}^{(b)}(x) \bmod x^{s(n-\tau)+\listl(n-k)-b(n-1)},
\end{split}
\end{equation}
with $\deg \overline{B}^{(b)}(x) < \listl(n-k)-s\tau+b$, $ b \in \SETzero{s}
$.
\end{definition}
The explicit expression for $S^{(b,t)}_i$ is difficult to obtain. 
We claim that it will not be easier to compute $S^{(b,t)}_i$ with such a formula than by calculating
the power series expansion of $T^{(b,t)}(x) = \overline R(x)^{t-b}/ \overline
G(x)^{s-b}$, which is fast to compute by computer algebra
techniques.

Considering the high degree terms, we get
$\sum_{b=0}^{s-1}(s-b)n = \binom{s+1}{2}n$ homogeneous equations from~\refeq{eq_GSKE_Synd}, which can be written as:
\begin{equation} \label{eq_GSKE_full_final}
\begin{split}
\sum_{t=b}^{\listl} \sum_{i=0}^{N_t-1} Q_{i}^{(t)} \cdot S^{(b,t)}_{j+i} = 0,  \quad & j \in \SETzero{(s-b)n},\\ 
& b \in \SETzero{s}.
\end{split}
\end{equation}
These linear equations lead to a Block-Hankel matrix. The
syndrome matrix $\mat{S} =\vert \vert \mat{S}^{(t,b)} \vert \vert$ for all $t \in \SETzero{\listl+1}, b \in \SETzero{s}$ of the reformulated Guruswami--Sudan
interpolation problem has the following form:
\begin{equation} \label{eq_GSKE_mat_full}
\left(\begin{array}{cccccc}
\mat{S}^{(0,0)} & \mat{S}^{(0,1)} & \dots & \dots & \dots & \mat{S}^{(0,\listl)} \\
 \mat{0}      & \mat{S}^{(1,1)} & \dots & \dots & \dots & \mat{S}^{(1,\listl)} \\
\vdots  & & \ddots & & \vdots \\
\mat{0} & \dots & \mat{0} & \mat{S}^{(s-1,s-1)} & \dots & \mat{S}^{(s-1,\listl)}
\end{array}\right),
\end{equation}
where each sub--matrix $\mat{S}^{(b,t)}= \vert \vert S^{(b,t)}_{i,j} \vert \vert $ 
is an $n(s-b) \times N_t$ Hankel matrix and $S^{(b,t)}(x) =
\sum_{i=0}^{(s-b)n+N_t-1} S^{(t,b)}_i x^i $ are the associated
polynomials with $S^{(b,t)}_{i,j} = S^{(b,t)}_{i+j}$.  All matrices
depend on the received vector $\mathbf{r}$ except the ones on the
diagonal: $\mat{S}^{(i,i)}$, $ i \in \SETzero{s} $.

\subsection{The FIA for the Block-Hankel Matrix of the Reformulated Guruswami--Sudan Interpolation Problem} \label{sec_BlockHankel}

We adapt the FIA to the Block-Hankel matrix of~\refeq{eq_GSKE_mat_full}. The structure of this syndrome
matrix is a mixture of the syndrome matrix (see Definition~\ref{def_SystemHankelHorizontalRearranged}) of the
reformulated Sudan interpolation problem and a vertical arrangement of
many Hankel matrices. The extension of the FIA for this case was
hinted in~\cite[Section 5.2]{Ruckenstein_PHD2001}.
First of all, let us express the $s$ Key Equations of~\refeq{eq_GSKE_full_final} in
terms of the inner product of bivariate polynomials.
\begin{problem}[Reformulated Guruswami--Sudan Problem] \label{prob_RearrangedBlockHankelSystem}
Let $S^{(b)}(x,y), \; b \in \SETzero{s}$ be $s$ bivariate
syndrome polynomials with:
\begin{equation} \label{eq_bivsyndGS}
 S^{(b)}(x,y) = \sum_{t=b}^{\listl} \sum_{i=0}^{N_t+(s-b)n-1} S^{(b,t)}_{i} x^i y^t,
\end{equation}
where the coefficients $S^{(b,t)}_{i}$ are given in Definition~\ref{def_syndGSwithQ0}. We search a nonzero bivariate
polynomial $T(x,y)$ that fulfills:
\begin{equation}
 \begin{split}
 \IP{x^{\kappa} T(x,y)}{S^{(\vartheta)}(x,y)} = 0,  \quad & \vartheta \in \SETzero{s},\\
& \kappa \in \SETzero{(s-\vartheta)n}.
 \end{split}
\end{equation}
\end{problem}

\printalgo{
\SetVline
\linesnumbered
\SetKwInput{KwData}{\underline{Data structures}}
\SetKwInput{KwResult}{\underline{Initialize}}
\SetKwInput{KwIn}{\underline{Input}}
\SetKwInput{KwOut}{\underline{Output}}
\KwIn{$s$ bivariate polynomials $S^{(b)}(x,y) = \sum_{t=0}^{\listl}S^{(t,b)}(x)y^t, b \in \SETzero{s}$;}
\KwOut{Bivariate polynomial $T(x,y)$;}
\BlankLine
\KwData{\\
Bivariate polynomial $T(x,y) = \sum_{t=0}^{\listl} T^{(t)}(x)y^{t}$; \\
Column pointers $(\nu,\mu)$, where $ \nu \in \SETzero{\listl}$, $\mu \in \SETzero{N_{\nu}}$;\\
Row pointer $(\vartheta,\kappa)$, where $\vartheta \in \SETzero{s}$ and $\kappa \in \SETzero{(s-\vartheta)n}$;\\
Two--dimensional array $A[(i,j)]$ of $\binom{s+1}{2} n $ entries in
$\Fxy$ indexed with the row pointer $(\vartheta,\kappa)$; \\
Two--dimensional array $D[(i,j)]$ of $\binom{s+1}{2} n $ entries in
$\F$ indexed with the row pointer $(\vartheta,\kappa)$;\\
Array $R$ of $\listl+1 $ entries containing the row pointer $(\vartheta,\kappa)$;\\
Variable $\Delta \in \F$, variable \textit{compute} $\in \lbrace \mathrm{TRUE, FALSE} \rbrace$;
}
\BlankLine
\KwResult{\\
Initialize arrays A, D and C to zero;\\
$(\nu,\mu) \leftarrow (0,0)$ and $(\vartheta, \kappa) \leftarrow (0,0) $; \\
\textit{compute} $ \leftarrow \mathrm{FALSE} $;
}
\BlankLine
\While{$(\vartheta,\kappa) < (s,0)$}{
\eIf{\textit{compute} }{$\Delta \leftarrow \langle x^{\kappa}\cdot T(x,y), S^{(\vartheta)}(x,y) \rangle $;\nllabel{alg_GSKE_computeDelta}}
{
  \eIf{$R[\nu] < 1$}
  {
  $ T(x,y) \leftarrow y^{\nu} \cdot x^{\mu} $\;
  $ \Delta \leftarrow S^{(0,\nu)}_{\mu}$\;
  $ (\vartheta, \kappa) \leftarrow (0,0)$\;
  }
  {
  $ T(x,y) \leftarrow x \cdot A[R[\nu]](x,y)$\;
  $ \Delta \leftarrow D[R[\nu]]$\;
  $ (\vartheta, \kappa) \leftarrow R[\nu]$\;

    \If{$\kappa  = 0$} 
    {
    $ (\vartheta, \kappa) \leftarrow (\vartheta - 1, n)$\;
    $ \Delta \leftarrow 0$\; 
    }
    $\kappa \leftarrow \kappa - 1$\; \nllabel{alg_GSKE_jump1}
  }
  $\text{\textit{compute}} \leftarrow \mathrm{TRUE}$\;
}
\eIf {$\Delta = 0$ or $D[(\vartheta,\kappa)] \neq 0$ \nllabel{alg_GSKE_jumping}} 
{ 
 \If{$\Delta \neq 0 $}{$T(x,y) \leftarrow T(x,y) - \frac{\Delta}{D[(\vartheta,\kappa)]} \cdot A[(\vartheta,\kappa)](x,y)$\; \nllabel{alg_GSKE_update}}
$ (\vartheta,\kappa) \leftarrow \nextv{\vartheta}{\kappa} $\; \nllabel{alg_GSKE_increment_row}
}(\textit{ /* $\Delta \neq 0$ and $D[(\vartheta,\kappa)] = 0$  */})
{
$ A[(\vartheta,\kappa)] \leftarrow T(x,y)  $\;
$ D[(\vartheta,\kappa)] \leftarrow \Delta$\;
$ R[\nu] \leftarrow (\vartheta,\kappa)$\; 
$ \text{\textit{compute}} \leftarrow \mathrm{FALSE}$\;
$ (\nu,\mu) \leftarrow \nexth{\nu}{\mu} $\; \nllabel{alg_GSKE_increment_column} 
}
}
\caption{Algorithm Solving Problem~\ref{prob_RearrangedBlockHankelSystem}}
\label{FIA_GSKE}
}

We adjust the FIA as an algorithm on a row- and column-interleaved version of the Block-Hankel matrix $\mat{S}$ of~\refeq{eq_GSKE_mat_full}.
Let us first define an ordering to describe the vertical rearrangement of the rows of the syndrome matrix $\mat{S}$ as in~\refeq{eq_GSKE_mat_full}.
Let denote $\prec_V$ the ordering on the rows, indexed by pairs $(\vartheta,\kappa)$, such that:
\begin{equation} \label{eq_prec}
\begin{split}
(\vartheta,\kappa) \prec_V & (\overline \vartheta, \overline \kappa) \iff \\
& \left\{\begin{array}{l}
\kappa+\vartheta n < \overline \kappa + \overline \vartheta n \\
\mbox{or} \\
\kappa+\vartheta n = \overline \kappa + \overline \vartheta n \mbox{ and } \vartheta < \overline \vartheta.
\end{array}\right.
\end{split}
\end{equation}
Let \nextv{\vartheta}{\kappa} denote the pair that immediately follows
$(\vartheta,\kappa)$ with respect to order defined by $\prec_V$ and let \prevv{\vartheta}{\kappa} denote the pair that immediately precedes
$(\vartheta,\kappa)$ with respect to order defined by $\prec_V$.
Furthermore, let:
\begin{equation} \label{eq_rowcounter}
R_{\vartheta,\kappa} = \big| \{ (t,i) \, | \, (t,i) \prec_V (\vartheta,\kappa)  \} \big|,
\end{equation}
which we use to index the rows of the virtually rearranged matrix
(similar to the horizontal case).
Note that $R_{\prevv{\vartheta}{\kappa}} = R_{\vartheta,\kappa}-1 $.

In the following, $\mat{S}^{\prime}$ denotes the rearranged version of the matrix $\mat{S}$ of~\refeq{eq_GSKE_mat_full}, where the
columns are ordered under $\prec_H$- and the rows under $\prec_V$-ordering.

Algorithm~\ref{FIA_GSKE} is the Fundamental Iterative Algorithm
tailored to a Block-Hankel matrix as in~\refeq{eq_GSKE_mat_full}.
As in the case of the reformulated Sudan interpolation problem, the columns of
the Block-Hankel matrix $\mat{S}$ are indexed by a couple $(\nu,
\mu)$, where $\nu \in \SETzero{\listl+1}$ and $\mu \in \SETzero{N_{\nu}}$.
Furthermore, the rows are indexed by a couple $(\vartheta, \kappa)$,
where $\vartheta \in \SETzero{s}$ and $\kappa \in \SETzero{(s-\vartheta)\cdot n}$.

Now, the arrays storing the discrepancies and the intermediate
polynomials are still indexed by rows, but the indexes of the rows
are two-dimensional, leading to two-dimensional arrays. The
two-dimensional array $A$ stores the intermediate bivariate
polynomials and the two-dimensional array $D$, stores the discrepancy
values. Both arrays $A$ and $D$ are indexed by the row pointer $(\vartheta,
\kappa)$. The discrepancy calculation (see Line~\ref{alg_GSKE_update}
of Algorithm~\ref{FIA_GSKE}) is adjusted to a Block-Hankel
matrix where each sub-horizontal band of Hankel matrices is
represented by a bivariate polynomial.

The intermediate bivariate connection polynomial $T^{(\vartheta,\kappa)}(x)$ of
Algorithm~\ref{FIA_GSKE} examining the $\kappa$th row and the $\mu$th column of the 
$(\nu, \vartheta)$th sub-matrix $\mat{S}^{(\nu,\vartheta)}$, gives us the vanishing linear combination of
the sub-matrix consisting of the first $R_{\vartheta, \kappa}$ rows
and the first $C_{\nu, \mu}$ columns of the rearranged syndrome matrix
$\mat{S}^{\prime}$. 

The row pointer of the sub-block $\vert \vert \mat{S}^{(\nu,0)} \, \mat{S}^{(\nu,1)} \, \cdots \, \mat{S}^{(\nu,s-1)} \vert \vert^T$ is stored in the array $R[\nu]$.
Note that $\listl+1$ row pointers of the form $(\vartheta, \kappa)$ need to be stored.

The adjusted initialization rule of Algorithm~\ref{FIA_GSKE} examining the Block-Hankel syndrome matrix as defined in~\refeq{eq_GSKE_mat_full} is stated in the following
proposition (see Line~\ref{alg_GSKE_jump1}, \ref{alg_GSKE_increment_row} and \ref{alg_GSKE_increment_column} of
Algorithm~\ref{FIA_GSKE}).
\begin{proposition}[Initialization Rule] \label{prop_FIA_BlockHankel}
Assume Algorithm~\ref{FIA_GSKE} examines column $(\nu,\mu)$ of a Block-Hankel syndrome matrix $\mat{S}$ as defined in
\refeq{eq_GSKE_mat_full} or equivalently the $s$ bivariate polynomials
$S^{(0)}(x,y),S^{(1)}(x,y),\dots,S^{(s-1)}(x,y)$ of Problem~\ref{prob_RearrangedBlockHankelSystem}. Assume that a
true discrepancy is obtained. Let $(\overline \nu, \overline \mu)= \nexth{\nu}{\mu}$ and let $(\vartheta,\kappa)$ be 
the previously stored value for the index of the last reached row in the sub-matrix of
index $\overline \nu$, and let $T(x,y)$ be the bivariate polynomial stored for
that row. If $({\overline \vartheta},{\overline \kappa}) = \prevv{\vartheta}{\kappa}$, we
can start examining column $ ({\overline \nu},{\overline \mu})$ of $\mat{S}$ at row
$(\overline \vartheta, \overline \kappa )$ with the initial value $\overline{T}(x,y) = x \cdot
T(x,y)$.
\end{proposition}
\begin{IEEEproof} \label{proof_FIA_BlockHankel}
In terms of the inner product (see Definition~\ref{def_InnerProduct}), we have:
\begin{equation} \label{eq:zerokappatheta}
\IP{x^{i_1} T(x,y)}{S^{(i_2)}(x,y)} = 0, \quad \forall \, (i_2,i_1) \prec_V (\overline \vartheta,\overline \kappa).
\end{equation}
Let us write $T(x,y)=\sum_{t=0}^{\overline \nu} \sum_{j=0}^{\overline \mu}  T^{(t)}_{j} x^j y^t$
and $\overline{T}(x,y)=\sum_{t=0}^{\overline \nu} \sum_{j=0}^{\overline \mu+1} \overline{T}^{(t)}_jx^j y ^t$, with $\overline{T}^{(t)}_j=T^{(t)}_{j-1}$, for $j>0$, and $\overline{T}^{(t)}_0=0$.  
Due to the structure of the Block-Hankel matrix $\mat{S}$, we have
the following identities:
\begin{align*}
 \IP{x^{i_1} \overline{T}(x,y)}{&S^{(i_2)}(x,y)} \\
 & = \sum_{t=0}^{\overline \nu} \sum_{j=1}^{\overline \mu+1} \overline{T}^{(t)}_j \cdot S^{(i_2,t)}_{i_1,j} \\
 & = \sum_{t=0}^{\overline \nu} \sum_{j=1}^{\overline \mu+1} T^{(t)}_{j-1} \cdot S^{(i_2,t)}_{i_1,j} \\
 & = \sum_{t=0}^{\overline \nu} \sum_{j=0}^{\overline \mu} T^{(t)}_j \cdot S^{(i_2,t)}_{i_1,j+1} \\
 & = \sum_{t=0}^{\overline \nu} \sum_{j=0}^{\overline \mu} T^{(t)}_j \cdot S^{(i_2,t)}_{i_1+1,j} \quad \textit{(Hankel)}\\
 & = \IP{x^{i_1+1} T(x,y)}{S^{(i_2)}(x,y)}
\end{align*}
which is zero for every $(i_2,i_1) \prec_V ({\overline \vartheta},{\overline \kappa})$.
\end{IEEEproof}

\begin{theorem}[Algorithm~\ref{FIA_GSKE}] 
Let $\mat{S}$ be the $\binom{s+1}{2}n \times \sum_{t=0}^{\listl} N_t$ syndrome Block-Hankel matrix of the
reformulated Guruswami--Sudan interpolation problem as in~\refeq{eq_GSKE_mat_full} and let
$S^{(b)}(x,y), b \in \SETzero{s}$ be the corresponding bivariate
syndrome polynomials as defined in
Problem~\ref{prob_RearrangedBlockHankelSystem}. Then
Algorithm~\ref{FIA_GSKE} outputs a bivariate polynomial $T(x,y)$, such
that:
\begin{equation*}
\begin{split}
\IP{x^{\kappa} T(x,y)}{S^{(\vartheta)}(x,y)} = 0,  \quad & \vartheta \in \SETzero{s},\\
 &\kappa \in \SETzero{(s-\vartheta)n}.
\end{split}
\end{equation*}
The time complexity of Algorithm~\ref{FIA_GSKE} is $\ON{\listl s^4 n^2}$.
\end{theorem}
\begin{IEEEproof}
The correctness is as usual, considering that we deal with the row- and
column-permuted version $\mat{S}^{\prime}$ of the Block-Hankel matrix $\mat{S}$ and that the initialization rule
is correct.

In the following, we analyze the complexity of Algorithm~\ref{FIA_GSKE}.
As in Section~\ref{sec_horizontalline}, we describe the state of Algorithm~\ref{FIA_GSKE} with the following triple:
\begin{equation}
\big( (\nu,\mu), ( [\vartheta,\kappa ]_0, [\vartheta,\kappa ]_1, \dots, [ \vartheta,\kappa ]_{\listl} ),  \delta \big),
\end{equation}
where $(\nu,\mu)$ is the current column pointer of Algorithm~\ref{FIA_GSKE},
when examining the $\mu$th column of the horizontal band of $s$ vertically
arranged Hankel matrices $\vert \vert \mat{S}^{(\nu,0)} \, \mat{S}^{(\nu,1)} \, \cdots \, \mat{S}^{(\nu,s-1)} \vert \vert^T $.
The index $[\vartheta,\kappa ]_{\nu}$ is the last considered
row in the horizontal band of $s$ sub-matrices $\vert \vert \mat{S}^{(\nu, 0)} \, \mat{S}^{(\nu, 1)} \, \cdots \, \mat{S}^{(\nu, s-1)} \vert \vert^T$.
These values are stored in the array $R$ of Algorithm~\ref{FIA_GSKE}.
As for Algorithm~\ref{FIA_RR}, $\delta$ denotes the number of already
encountered true discrepancies.
Assume $(\nu,\mu)$ is the current column pointer of Algorithm~\ref{FIA_GSKE}.
The same two cases as before can happen:\\
1) Either, there is no true discrepancy, then Algorithm~\ref{FIA_GSKE} remains
in the same column $(\nu,\mu)$ of the sub-matrices $\vert \vert \mat{S}^{(\nu, 0)} \, \mat{S}^{(\nu, 1)} \, \cdots \, \mat{S}^{(\nu, s-1)} \vert \vert^T $
and the triple becomes:
\begin{equation*}
\begin{split}
\big( (\nu,\mu), & ( [\vartheta,\kappa ]_0, [\vartheta,\kappa ]_1,  \dots, \\
 & [\vartheta,\kappa ]_\nu \leftarrow \mathsf{next}\lbrack \prec_V,( [\vartheta,\kappa ]_\nu) \rbrack , \dots , [ \vartheta,\kappa ]_{\listl} ),  \delta \big),
\end{split} 
\end{equation*}
2) Or, a true discrepancy is encountered and the triple becomes:
\begin{equation*}
\begin{split}
\big( (\overline \nu,\overline \mu), & ( [\vartheta,\kappa ]_0, [\vartheta,\kappa ]_1, \dots, \\
 & [\vartheta,\kappa ]_{\overline \nu} \leftarrow \mathsf{prev}\lbrack \prec_V,( [\vartheta,\kappa ]_{\overline \nu}) \rbrack , \dots , [ \vartheta,\kappa ]_{\listl} ),  \delta +1 \big),
\end{split}
\end{equation*}
where $(\overline \nu, \overline \mu) \gets \nexth{\nu}{\mu}$.
In both cases, the sum $\mathsf{Iter}$ of the triple is:
\begin{equation}
\mathsf{Iter} =  C_{\nu,\mu} + \left( \sum_{t \in \SETzero{\listl+1}} R_{[\vartheta,\kappa ]_t} \right) + \delta,
\end{equation}
when Algorithm~\ref{FIA_GSKE} examines the $(\nu,\mu)$th column of the Block-Hankel matrix $\mat{S}$ of~\refeq{eq_GSKE_mat_full} and 
it increases by one in each iteration.
The initial value of $\mathsf{Iter}$ is zero, and the
final value can be bounded by
\begin{align*}
\mathsf{Iter} & \leq  \binom{s+1}2n + \sum_{t=0}^{\listl} \binom{s+1}2n  + \binom{s+1}2n\\
& \leq \ON{\listl s^2n}.
\end{align*}
The number of iterations of Algorithm~\ref{FIA_GSKE} is bounded by \ON{\listl s^2n}.

This gives a total of $\ON{\listl s^4 n^2}$, since the discrepancy calculation requires $\ON{s^2n}$.
\end{IEEEproof}

\subsection{Example: Guruswami--Sudan Decoding of a Generalized Reed--Solomon Code with Adapted FIA} \label{subsec_GSKE}
We consider the case of multiplicity $s=2$ for the \GRS{16}{4} code. 
The corresponding list size is $\listl=4$. The decoding radius is now $\tau=8$. The
degrees of the univariate polynomials $Q^{(t)}(x)$ are
$(N_0,N_1,N_2,N_3,N_4) = (16,13,10,7,4)$. 

The Block-Hankel syndrome matrix \mat{S}
\begin{equation*}
 \mat{S} = \left( \begin{array}{ccccc}
\mat{S}^{(0,0)} & \mat{S}^{(0,1)} & \mat{S}^{(0,2)} & \mat{S}^{(0,3)} & \mat{S}^{(0,4)} \\
\mat{0} & \mat{S}^{(1,1)} & \mat{S}^{(1,2)} & \mat{S}^{(1,3)} & \mat{S}^{(1,4)} \\
\end{array}\right)
\end{equation*}
is a $(3n=48) \times (\sum_{t=0}^{4} N_t = 50)$ matrix. 
It consists of nine non-zero Hankel matrices and one all-zero matrix $\mat{S}^{(1,0)}$ arranged in two horizontal
bands of five Hankel matrices.
\begin{figure*}[htb]
    \centering
       \includegraphics[scale=1]{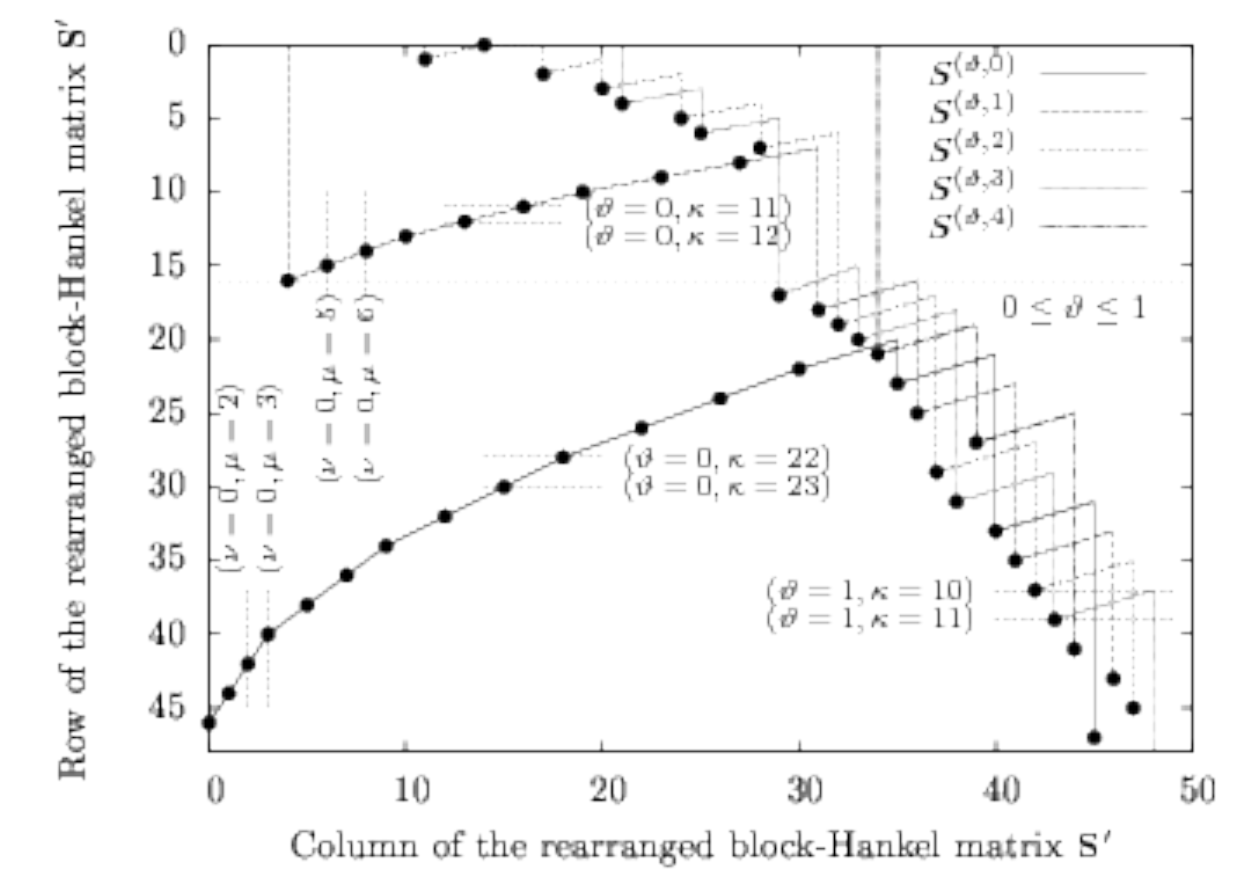}
        \caption{Illustration of the row pointer $(\vartheta,\kappa)$
          of Algorithm~\ref{FIA_GSKE} applied to a $48 \times 50$
          Block-Hankel matrix $\mat{S}$. The matrix consists of two
          vertically arranged bands of five horizontally arranged
          Hankel matrices.  The first band consists of 32 rows and the
          second one of 16. The plotted matrix $\mat{S}^{\prime}$ consists
          of the rearranged columns and rows of the matrix $\mat{S}$
          under $\prec_H$- respective $\prec_V$-ordering . The mixture
          of rows of the two vertical lines starts in line 16 (marked
          by the dotted horizontal line). The five zig--zag lines
          $S^{(\vartheta,0)},S^{(\vartheta,1)}, \dots
          ,S^{(\vartheta,4)}$ trace the row pointer for the five
          sub-blocks $\vert \vert \mat{S}^{(0,0)} \, \mat{S}^{(1,0)} \vert \vert^T$, $\vert \vert \mat{S}^{(1,0)} \, \mat{S}^{(1,1)} \vert \vert^T$, $\dots$, 
          $\vert \vert \mat{S}^{(4,0)} \, \mat{S}^{(4,1)} \vert \vert^T$ of two vertically arranged Hankel matrices.}
        \label{fig_ExampleGSKE}
\end{figure*}
The values of the row pointer $(\vartheta,\kappa)$ of Algorithm~\ref{FIA_GSKE} for the Block-Hankel matrix are traced in Figure~\ref{fig_ExampleGSKE}. 
The five zig--zag lines $S^{(\vartheta,0)},S^{(\vartheta,1)}, \dots
,S^{(\vartheta,4)}$ in Figure~\ref{fig_ExampleGSKE} trace the row pointer $(\vartheta,\kappa)$, when Algorithm~\ref{FIA_GSKE} 
examines the five sub-blocks $\vert \vert \mat{S}^{(0,0)} \, \mat{S}^{(1,0)} \vert \vert^T$, $\vert \vert \mat{S}^{(0,1)} \, \mat{S}^{(1,1)} \vert \vert^T$, $\dots$,
$\vert \vert \mat{S}^{(0,4)} \, \mat{S}^{(1,4)} \vert \vert^T$. 
In Table~\ref{tb_column pointer_GSKE}, the column $C_{\nu,\mu}$ and
the column pointer $(\nu,\mu)$ according to $\prec_{H}$ for the syndrome matrix of the \GRS{16}{4} code are listed.
Additionally to the horizontal ordering $\prec_{H}$ of the columns (as in the Sudan case), now the rows are ordered according to $\prec_{V}$.
The row-index $R_{\vartheta,\kappa}$ and the row pointer $(\vartheta,
\kappa)$ are shown in Table~\ref{tb_row pointer_GSKE}.
\begin{table}[htb]
      \renewcommand{\arraystretch}{1.2}
      \centering
       \caption{Column-index $C_{\nu,\mu}$ and column pointer $(\nu,\mu)$ for the Block-Hankel syndrome matrix of a $\mathcal{GRS}(16,4)$ code with multiplicity $s=2$ and list size $\listl=4$.}
      \begin{tabular}{cccccccccc}
      \hline
      \multicolumn{10}{c}{\textbf{Column $C_{\nu,\mu}$ and }}  \\
      \multicolumn{10}{c}{\textbf{Column pointer $(\nu,\mu)$}}  \\
      \hline\hline
  0 & (0,0) & 10 & (1,3) & 20 & (2,3)  & 30 & (0,12) & 40 & (0,14) \\
  1 & (0,1) & 11 & (2,0) & 21 & (3,0)  & 31 & (1,9)  & 41 & (1,11) \\
  2 & (0,2) & 12 & (0,7) & 22 & (0,10) & 32 & (2,6)  & 42 & (2,8)  \\
  3 & (0,3) & 13 & (1,4) & 23 & (1,7)  & 33 & (3,3)  & 43 & (3,5)  \\
  4 & (1,0) & 14 & (2,1) & 24 & (2,4)  & 34 & (4,0)  & 44 & (4,2)  \\
  5 & (0,4) & 15 & (0,8) & 25 & (3,1)  & 35 & (0,13) & 45 & (0,15) \\
  6 & (1,1) & 16 & (1,5) & 26 & (0,11) & 36 & (1,10) & 46 & (1,12) \\
  7 & (0,5) & 17 & (2,2) & 27 & (1,8)  & 37 & (2,7)  & 47 & (2,9)  \\
  8 & (1,2) & 18 & (0,9) & 28 & (2,5)  & 38 & (3,4)  & 48 & (3,6)  \\
  9 & (0,6) & 19 & (1,6) & 29 & (3,2)  & 39 & (4,1)  & 49 & (4,3) \\
      \hline
      \end{tabular}
      \label{tb_column pointer_GSKE}
\end{table}
Let us consider three cases, where a true discrepancy in Algorithm~\ref{FIA_GSKE} occurred.
The first case are the most left two points in Figure~\ref{fig_ExampleGSKE}.
The value of the column pointer $(\nu,\mu)$ is $(0,2)$ and $(0,3)$.
Algorithm~\ref{fig_ExampleGSKE} examines the first band of the two
Hankel matrices $\vert \vert \mat{S}^{(0,0)} \, \mat{S}^{(1,0)} \vert \vert^T$
traced by line $S^{(\vartheta,0)}$. For the first
pair no columns were virtually interchanged and the horizontal
distance is one.

The second two points with the values of the column pointer $(0,5)$ and $(0,6)$ indicate a
true discrepancy of Algorithm~\ref{fig_ExampleGSKE}, when the second band of the two
Hankel matrices $\vert \vert \mat{S}^{(0,1)} \, \mat{S}^{(1,1)} \vert \vert^T$ is
examined. The values are traced by the line $S^{(\vartheta,1)}$ in Figure~\ref{fig_ExampleGSKE}.  
For the second pair ($(0,5)$,$(0,6)$), the columns of the 
first and second vertical band of Hankel
matrices are mixed and therefore the horizontal distance is two.

\begin{table}[htb]
      \renewcommand{\arraystretch}{1.2}
      \centering
       \caption{Row-index $R_{\vartheta,\kappa}$ and row pointer $(\vartheta, \kappa)$ of Algorithm~\ref{FIA_GSKE} for
         Block-Hankel syndrome matrix of a $\mathcal{GRS}(16,4)$ code with multiplicity $s=2$ and list size $\listl=4$.}
      \begin{tabular}{cc}
      \hline
      \multicolumn{2}{c}{\textbf{Row $R_{\vartheta,\kappa}$ and }}  \\
      \multicolumn{2}{c}{\textbf{Row pointer $(\vartheta,\kappa)$}}  \\
      \hline\hline
  0 - 16 & (0,0 - 16) \\
  17 & (1,0) \\
  18 & (0,17) \\
  19 & (1,1) \\
  20 & (0,18) \\
  $\vdots$ & $\vdots$ \\
  46 & (0,31) \\
  47 & (1,15) \\
      \hline
      \end{tabular}
      \label{tb_row pointer_GSKE}
\end{table}
The third considered case, where a true discrepancy occurs, are the most right two points in Figure~\ref{fig_ExampleGSKE}
indicated by values $(1,10)$ and $(1,11)$ of the row pointer $(\vartheta,\kappa)$.
Algorithm~\ref{FIA_GSKE} examines the band of the two
Hankel matrices $\vert \vert \mat{S}^{(0,3)} \, \mat{S}^{(1,3)} \vert \vert^T$
and restarts (at the point $(1,10)$) with the previous stored
value of the row pointer (at $(1,11)$). In between four other
horizontal bands of matrices were examined.


\section{Conclusion} \label{sec_conclusion}
We reformulated the Guruswami--Sudan interpolation conditions (for a
multiplicity higher than one) for Generalized Reed--Solomon codes into
a set of univariate polynomial equations, which can partially be seen
as Extended Key Equations. The obtained set of homogeneous linear equations has
a Block-Hankel structure. We adapted the Fundamental Iterative
Algorithm of Feng and Tzeng to this special structure and achieved a
significant reduction of the time complexity.

As mentioned in Note~\ref{note:GSDiagonal}, the set of equations can be
further reduced, under the observation that the diagonal terms are
constant, i.e., they do not depend on the received word. This reduction
leads to a loss of the Block-Hankel structure and therefore would
destroy the quadratic complexity. We note that Beelen and
H{\o}holdt~\cite{BeelenHoholdt_TheDecodingofAGCodes_2008} mentioned
this reduction for the Guruswami--Sudan interpolation step for Algebraic Geometric
codes, to get a smaller interpolation problem, but the system does not
appear to be Block-Hankel.

We conclude that we identified the quantity
$\lfloor (s\tau)/d \rfloor$ (see Note~\ref{note:GSDiagonal})
without having found an interpretation of that number.

\section{Acknowledgment}
The authors thank Vladimir Sidorenko and Martin Bossert for fruitful discussions.

We thank the anonymous referees for their valuable comments that improved 
the presentation of this paper.

\IEEEtriggeratref{31}



\textbf{Alexander Zeh} studied electrical engineering at the University of Applied Science in Stuttgart, with the main topic automation technology.
He received his Dipl.-Ing. (BA) degree in 2004. He continued his studies at Universität Stuttgart until 2008, where he received
is Dipl.-Ing. in electrical engineering. He participated in the double-diploma program with Télécom ParisTech (former ENST) from
2006 to 2008 and he received also a french diploma.
Currently he is a Ph.D. student at the Institute of Telecommunications and Applied Information Theory, University of Ulm, Germany
and at the Computer Science Department (LIX), École Polytechnique ParisTech, Paris, France.
His current research interests include coding and information theory, signal processing, telecommunications and the implementation of fast algorithms on FPGAs.
\vspace{1ex}

\textbf{Christian Gentner} studied electrical engineering at the University of Applied Science in Ravensburg, with the main topic communication technology.
He received his Dipl.-Ing. (BA) degree in 2006.
He continued his studies at the University of Ulm until 2009, where he received his M.Sc. in telecommunication and media technology.
He is currently working towards the Ph.D. degree at the Institute of Communications and Navigation of the German Aerospace Center (DLR), Germany. His current research interests include multi-sensor navigation, propagation effects and non-line-of-sight identification and mitigation as well as the implementation of these algorithms on FPGAs.

\vspace{1ex}

\textbf{Daniel Augot} studied Mathematics and Computer Science at the
University Pierre and Marie Curie in Paris. He received the Master
degree in theoretical computer science in 1989. He was a Ph.D. student
of Pascale Charpin and graduated in 1993. He was then hired as a
researcher at INRIA--Rocquencourt. In 2009, he became a senior
researcher at INRIA--Saclay-Île-de-France and École Polytechnique. His
major research interests are coding theory, cryptography and their
interplay.

\vspace{1ex}


\begin{thebibliography}{10}
\providecommand{\url}[1]{#1}
\csname url@samestyle\endcsname
\providecommand{\newblock}{\relax}
\providecommand{\bibinfo}[2]{#2}
\providecommand{\BIBentrySTDinterwordspacing}{\spaceskip=0pt\relax}
\providecommand{\BIBentryALTinterwordstretchfactor}{4}
\providecommand{\BIBentryALTinterwordspacing}{\spaceskip=\fontdimen2\font plus
\BIBentryALTinterwordstretchfactor\fontdimen3\font minus
  \fontdimen4\font\relax}
\providecommand{\BIBforeignlanguage}[2]{{%
\expandafter\ifx\csname l@#1\endcsname\relax
\typeout{** WARNING: IEEEtran.bst: No hyphenation pattern has been}%
\typeout{** loaded for the language `#1'. Using the pattern for}%
\typeout{** the default language instead.}%
\else
\language=\csname l@#1\endcsname
\fi
#2}}
\providecommand{\BIBdecl}{\relax}
\BIBdecl

\bibitem{AugotZeh_OnTheRREquations_2008}
\BIBentryALTinterwordspacing
D.~Augot and A.~Zeh, ``On the {R}oth and {R}uckenstein {E}quations for the
  {G}uruswami-{S}udan {A}lgorithm,'' in \emph{Information Theory, 2008. ISIT
  2008. IEEE International Symposium on}, 2008, pp. 2620--2624. [Online].
  Available: \url{http://dx.doi.org/10.1109/ISIT.2008.4595466}
\BIBentrySTDinterwordspacing

\bibitem{Zeh_EfficientListDecoding_2009}
\BIBentryALTinterwordspacing
A.~Zeh, C.~Gentner, and M.~Bossert, ``Efficient {L}ist-{D}ecoding of
  {R}eed-{S}olomon {C}odes with the {F}undamental {I}terative {A}lgorithm,'' in
  \emph{Information Theory Workshop, 2009. ITW 2009. IEEE}, October 2009.
  [Online]. Available: \url{http://dx.doi.org/10.1109/ITW.2009.5351241}
\BIBentrySTDinterwordspacing

\bibitem{GuruswamiSudan_ImproveddecodingofReed-Solomonandalgebraic-geometrycod%
es_1999}
\BIBentryALTinterwordspacing
V.~Guruswami and M.~Sudan, ``{Improved Decoding of Reed-Solomon and
  Algebraic-Geometry Codes},'' \emph{IEEE Transactions on Information Theory},
  vol.~45, no.~6, pp. 1757--1767, 1999. [Online]. Available:
  \url{http://ieeexplore.ieee.org/xpls/abs\_all.jsp?arnumber=782097}
\BIBentrySTDinterwordspacing

\bibitem{Guruswami_ListDecodingofError-CorrectingCodes_1999}
V.~Guruswami, \emph{List Decoding of Error-Correcting Codes}, ser. Lecture Notes in Computer Science.\hskip 1em plus 0.5em minus
  0.4em\relax New York: Springer, 2004, no. 3282.

\bibitem{Guruswami_ALGORITHMICRESULTSINLISTDECODING_2007}
------, \emph{Algorithmic Results in List Decoding}.\hskip 1em plus 0.5em minus
  0.4em\relax Now Publishers Inc, January 2007.

\bibitem{sudan97decoding}
\BIBentryALTinterwordspacing
M.~Sudan, ``Decoding of {R}eed-{S}olomon {C}odes beyond the
  {E}rror-{C}orrection {B}ound,'' \emph{Journal of Complexity}, vol.~13, no.~1,
  pp. 180--193, March 1997. [Online]. Available:
  \url{http://dx.doi.org/10.1006/jcom.1997.0439}
\BIBentrySTDinterwordspacing

\bibitem{ReedSolomon_PolynomialCodesOverCertainFiniteFields_1960}
\BIBentryALTinterwordspacing
I.~S. Reed and G.~Solomon, ``Polynomial {C}odes {O}ver {C}ertain {F}inite
  {F}ields,'' \emph{Journal of the Society for Industrial and Applied
  Mathematics}, vol.~8, no.~2, pp. 300--304, 1960. [Online]. Available:
  \url{http://dx.doi.org/10.1137/0108018}
\BIBentrySTDinterwordspacing

\bibitem{Koetter:THS1996}
R.~Koetter, ``{On Algebraic Decoding of Algebraic-Geometric and Cyclic
  Codes},'' Ph.D. dissertation, University of Link\"{o}ping, 1996.

\bibitem{Roth_Ruckenstein_2000}
\BIBentryALTinterwordspacing
R.~M. Roth and G.~Ruckenstein, ``{Efficient Decoding of Reed--Solomon Codes
  Beyond Half the Minimum Distance},'' \emph{Information Theory, IEEE
  Transactions on}, vol.~46, no.~1, pp. 246--257, 2000. [Online]. Available:
  \url{http://ieeexplore.ieee.org/xpls/abs\_all.jsp?arnumber=817522}
\BIBentrySTDinterwordspacing

\bibitem{Ruckenstein_PHD2001}
\BIBentryALTinterwordspacing
G.~Ruckenstein, ``{Error Decoding Strategies for Algebraic Codes},'' Ph.D.
  dissertation, Technion, 2001. [Online]. Available:
  \url{http://www.cs.technion.ac.il/users/wwwb/cgi-bin/tr-info.cgi/2001/PHD/PH%
D-2001-01}
\BIBentrySTDinterwordspacing

\bibitem{Alekhnovich_LinearDiophantineEquation_2005}
\BIBentryALTinterwordspacing
M.~Alekhnovich, ``{Linear Diophantine Equations Over Polynomials and Soft
  Decoding of Reed-Solomon Codes},'' \emph{Information Theory, IEEE
  Transactions on}, vol.~51, no.~7, pp. 2257--2265, 2005. [Online]. Available:
  \url{http://ieeexplore.ieee.org/xpls/abs\_all.jsp?arnumber=1459042}
\BIBentrySTDinterwordspacing

\bibitem{Trifonov_EfficientInterpolation_2010}
\BIBentryALTinterwordspacing
P.~Trifonov, ``{Efficient Interpolation in the Guruswami--Sudan Algorithm},''
  \emph{Information Theory, IEEE Transactions on}, vol.~56, no.~9, pp. 4341
  --4349, Sep. 2010. [Online]. Available:
  \url{http://dx.doi.org/10.1109/TIT.2010.2053901}
\BIBentrySTDinterwordspacing

\bibitem{Sakata_FindingAMinimalSet_1991}
\BIBentryALTinterwordspacing
S.~Sakata, ``Finding a minimal polynomial vector set of a vector of {nD}
  arrays,'' in \emph{Proceedings of Applied Algebra, Algebraic Algorithms and
  Error--Correcting Codes ({AAECC} '91)}, ser. LNCS, H.~F. Mattson, T.~Mora,
  and T.~R.~N. Rao, Eds., vol. 539.\hskip 1em plus 0.5em minus 0.4em\relax
  Berlin, Germany: Springer, Oct. 1991, pp. 414--425. [Online]. Available:
  \url{http://dx.doi.org/10.1007/3-540-54522-0\_129}
\BIBentrySTDinterwordspacing

\bibitem{Sakata_OnFastInterpolation_2001}
\BIBentryALTinterwordspacing
------, ``{On Fast Interpolation Method for Guruswami-Sudan List Decoding of
  One-Point Algebraic-Geometry Codes},'' in \emph{Applied Algebra, Algebraic
  Algorithms and Error-Correcting Codes}, ser. Lecture Notes in Computer
  Science, S.~Bozta\c{s} and I.~E. Shparlinski, Eds.\hskip 1em plus 0.5em minus
  0.4em\relax Berlin, Heidelberg: Springer, October 2001, vol. 2227, ch.~18,
  pp. 172--181. [Online]. Available:
  \url{http://dx.doi.org/10.1007/3-540-45624-4\_18}
\BIBentrySTDinterwordspacing

\bibitem{Feng_Tzeng1991}
\BIBentryALTinterwordspacing
G.~L. Feng and K.~K. Tzeng, ``{A Generalization of the Berlekamp-Massey
  Algorithm for Multisequence Shift-Register Synthesis with Applications to
  Decoding Cyclic Codes},'' \emph{Information Theory, IEEE Transactions on},
  vol.~37, no.~5, pp. 1274--1287, 1991. [Online]. Available:
  \url{http://ieeexplore.ieee.org/xpls/abs\_all.jsp?arnumber=133246}
\BIBentrySTDinterwordspacing

\bibitem{Berlekamp:AGT1968}
E.~R. Berlekamp, \emph{Algebraic coding theory}.\hskip 1em plus 0.5em minus
  0.4em\relax McGraw-Hill, 1968.

\bibitem{Massey_Shift-registersynthesisandBCHdecoding_1969}
\BIBentryALTinterwordspacing
J.~Massey, ``{Shift-Register Synthesis and BCH Decoding},'' \emph{Information
  Theory, IEEE Transactions on}, vol.~15, no.~1, pp. 122--127, January 2003.
  [Online]. Available:
  \url{http://ieeexplore.ieee.org/xpls/abs\_all.jsp?arnumber=1054260}
\BIBentrySTDinterwordspacing

\bibitem{Beelen_ASyndromeFormulation_2008}
\BIBentryALTinterwordspacing
P.~Beelen and T.~H{\o}holdt, ``{A Syndrome Formulation of the Interpolation
  Step in the Guruswami-Sudan Algorithm},'' in \emph{ICMCTA}, ser. Lecture
  Notes in Computer Science, A.~I. Barbero, Ed., vol. 5228.\hskip 1em plus
  0.5em minus 0.4em\relax Springer, 2008, pp. 20--32. [Online]. Available:
  \url{http://dx.doi.org/10.1007/978-3-540-87448-5_3}
\BIBentrySTDinterwordspacing

\bibitem{Beelen_KeyEquations_2010}
P.~Beelen and K.~Brander, ``{Key Equations for List Decoding of Reed--Solomon
  Codes and How to Solve Them},'' \emph{Journal of Symbolic Computation},
  vol.~45, no.~7, pp. 773--786, Jul. 2010.

\bibitem{BerlekampWelch_Patent}
E.~R. Berlekamp and L.~Welch, ``Error correction of algebraic block codes,'' US
  Patent Number 4,633,470.

\bibitem{MacWilliamsSloane_TheTheoryOfErrorCorrecting_1988}
F.~J. MacWilliams and N.~J.~A. Sloane, \emph{The Theory of Error-Correcting
  Codes (North-Holland Mathematical Library)}.\hskip 1em plus 0.5em minus
  0.4em\relax North Holland, June 1988.

\bibitem{Roth_IntroductiontoCodingTheory_2006}
R.~Roth, \emph{Introduction to Coding Theory}.\hskip 1em plus 0.5em minus
  0.4em\relax Cambridge University Press, March 2006.

\bibitem{Hasse_1936}
H.~Hasse, ``Theorie der h\"{o}heren {D}ifferentiale in einem algebraischen
  {F}unktionenk\"{o}rper mit vollkommenem {K}onstantenk\"{o}rper bei beliebiger
  {C}harakteristik.'' \emph{J. Reine Angew. Math.}, vol. 175, pp. 50--54, 1936.

\bibitem{Gemmell-Sudan_IFL1992}
P.~Gemmell and M.~Sudan, ``Highly resilient correctors for polynomials,''
  \emph{Information Processing Letters}, vol.~43, no.~4, pp. 169--174, Sep.
  1992.

\bibitem{YaghoobianBlake_TwoNewDecodingAlgorithmsForRSCodes_1994}
\BIBentryALTinterwordspacing
T.~Yaghoobian and I.~F. Blake, ``{Two new decoding algorithms for Reed-Solomon
  codes},'' \emph{Applicable Algebra in Engineering, Communication and
  Computing}, vol.~5, no.~1, pp. 23--43, January 1994. [Online]. Available:
  \url{http://dx.doi.org/10.1007/BF01196623}
\BIBentrySTDinterwordspacing

\bibitem{DabiriBlake_FastParallelAlgorithmsRemainder_1995}
\BIBentryALTinterwordspacing
D.~Dabiri and I.~F. Blake, ``Fast parallel algorithms for decoding
  {R}eed-{S}olomon codes based on remainder polynomials,'' \emph{Information
  Theory, IEEE Transactions on}, vol.~41, no.~4, pp. 873--885, 1995. [Online].
  Available: \url{http://dx.doi.org/10.1109/18.391235}
\BIBentrySTDinterwordspacing

\bibitem{JustesenHoholdt_ACourseinError-CorrectingCodes_2004}
J.~Justesen and T.~Hoholdt, \emph{A Course in Error-Correcting Codes (EMS
  Textbooks in Mathematics)}.\hskip 1em plus 0.5em minus 0.4em\relax European
  Mathematical Society, February 2004.

\bibitem{Sugiyama_AMethodOfSolving_1975}
Y.~Sugiyama, M.~Kasahara, S.~Hirasawa, and T.~Namekawa, ``{A Method for Solving
  Key Equation for Decoding Goppa Codes},'' \emph{Information and Control},
  vol.~27, no.~1, pp. 87--99, 1975.

\bibitem{Dornstetter_OnTheEquivalence_1987}
\BIBentryALTinterwordspacing
J.~Dornstetter, ``{On the Equivalence Between Berlekamp's and Euclid's
  Algorithms},'' \emph{Information Theory, IEEE Transactions on}, vol.~33,
  no.~3, pp. 428--431, 1987. [Online]. Available:
  \url{http://ieeexplore.ieee.org/xpls/abs\_all.jsp?arnumber=1057299}
\BIBentrySTDinterwordspacing

\bibitem{Heydtmann_BMEuclidFIA_2000}
\BIBentryALTinterwordspacing
A.~E. Heydtmann and J.~M. Jensen, ``{On the Equivalence of the Berlekamp-Massey
  and the Euclidean Algorithms for Decoding},'' \emph{Information Theory, IEEE
  Transactions on}, vol.~46, no.~7, pp. 2614--2624, 2000. [Online]. Available:
  \url{http://dx.doi.org/10.1109/18.887869}
\BIBentrySTDinterwordspacing

\bibitem{Amoros-Sullivan:AAECC2009}
\BIBentryALTinterwordspacing
M.~Bras-Amor{\'o}s and M.~E. O'Sullivan, ``From the Euclidean Algorithm for
  Solving a Key Equation for Dual Reed--Solomon Codes to the Berlekamp--Massey
  Algorithm,'' in \emph{AAECC}, ser. Lecture Notes in Computer Science,
  M.~Bras-Amor{\'o}s and T.~H{\o}holdt, Eds., vol. 5527.\hskip 1em plus 0.5em
  minus 0.4em\relax Springer, 2009, pp. 32--42. [Online]. Available:
  \url{http://dx.doi.org/10.1007/978-3-642-02181-7}
\BIBentrySTDinterwordspacing

\bibitem{Schmidt_CollaborativeDecoding_2009}
\BIBentryALTinterwordspacing
G.~Schmidt, V.~R. Sidorenko, and M.~Bossert, ``Collaborative {D}ecoding of
  {I}nterleaved {R}eed-{S}olomon {C}odes and {C}oncatenated {C}ode {D}esigns,''
  \emph{Information Theory, IEEE Transactions on}, vol.~55, no.~7, pp.
  2991--3012, 2009. [Online]. Available:
  \url{http://dx.doi.org/10.1109/TIT.2009.2021308}
\BIBentrySTDinterwordspacing

\bibitem{McEliece_TheGuruswami-SudanDecodingAlgorithmforReed-SolomonCodes_2003}
\BIBentryALTinterwordspacing
R.~J. Mceliece, ``The {G}uruswami--{S}udan {D}ecoding {A}lgorithm for
  {R}eed--{S}olomon {C}odes,'' \emph{Interplanetary Network Progress Report},
  vol. 153, pp. 1--60, January 2003. [Online]. Available:
  \url{http://ipnpr.jpl.nasa.gov/progress\_report/42-153/153F.pdf}
\BIBentrySTDinterwordspacing

\bibitem{BeelenHoholdt_TheDecodingofAGCodes_2008}
P.~Beelen and T.~Hoholdt, ``{The Decoding of Algebraic Geometry Codes},'' in
  \emph{Series on Coding Theory and Cryptology : Advances in Algebraic Geometry
  Codes ; volume 5}.\hskip 1em plus 0.5em minus 0.4em\relax World Scientific,
  2008, pp. 49--98.

\end{thebibliography}
\end{document}